\documentclass[preprint,12pt,authoryear,appendixfloats]{aastex61}

\usepackage{lineno}

\begin{document}

\title{Analysis of Neptune's 2017 Bright Equatorial Storm}

\author{Edward Molter}
\affiliation{Astronomy Department, University of California, Berkeley; Berkeley CA, 94720, USA}

\correspondingauthor{Edward Molter}
\email{emolter@berkeley.edu}

\author{Imke de Pater}
\affiliation{Astronomy Department, University of California, Berkeley; Berkeley CA, 94720, USA}
\affiliation{Earth and Planetary Science Department, University of California, Berkeley; Berkeley CA, 94720, USA}

\author{Statia Luszcz-Cook}
\affiliation{Department of Astronomy, Columbia University, Pupin Hall, 538 West 120th Street, New York City, NY 10027}
\affiliation{Astrophysics Department, American Museum of Natural History, Central Park West at 79th Street, New York, NY 10024, USA}

\author{Ricardo Hueso}
\affiliation{Departamento F{\'i}sica Aplicada I, Escuela Ingenier{\'i}a de Bilbao, Uiversidad del Pa{\'i}s Vasco UPV/EHU (Spain)}

\author{Joshua Tollefson}
\affiliation{Earth and Planetary Science Department, University of California, Berkeley; Berkeley CA, 94720, USA}

\author{Carlos Alvarez}
\affiliation{W. M. Keck Observatory, 65-1120 Mamalahoa Hwy., Kamuela, HI 96743, USA}

\author{Agust{\'i}n S{\'a}nchez-Lavega}
\affiliation{Departamento F{\'i}sica Aplicada I, Escuela Ingenier{\'i}a de Bilbao, Uiversidad del Pa{\'i}s Vasco UPV/EHU (Spain)}

\author{Michael H. Wong}
\affiliation{Astronomy Department, University of California, Berkeley; Berkeley CA, 94720, USA}

\author{Andrew I. Hsu}
\affiliation{Astronomy Department, University of California, Berkeley; Berkeley CA, 94720, USA}

\author{Lawrence A. Sromovsky}
\affiliation{University of Wisconsin, Madison, WI, USA}

\author{Patrick M. Fry}
\affiliation{University of Wisconsin, Madison, WI, USA}

\author{Marc Delcroix}
\affiliation{Planetary observations section, Soci{\'e}t{\'e} Astronomique de France, France}

\author{Randy Campbell}
\affiliation{W. M. Keck Observatory, 65-1120 Mamalahoa Hwy., Kamuela, HI 96743, USA}

\author{Katherine de Kleer}
\affiliation{Division of Geological and Planetary Sciences, California Institute of Technology, Pasadena, CA 91125}

\author{Elinor Gates}
\affiliation{Lick Observatory, PO Box 85, Mount Hamilton, CA 95140}

\author{Paul David Lynam}
\affiliation{Lick Observatory, PO Box 85, Mount Hamilton, CA 95140}

\author{S. Mark Ammons}
\affiliation{Lawrence Livermore National Laboratory, 7000 East Ave., Livermore, CA 94550}

\author{Brandon Park Coy}
\affiliation{Astronomy Department, University of California, Berkeley; Berkeley CA, 94720, USA}

\author{Gaspard Duchene}
\affiliation{Astronomy Department, University of California, Berkeley; Berkeley CA, 94720, USA}
\affiliation{Univ. Grenoble Alpes/CNRS, IPAG, F-38000 Grenoble, France}

\author{Erica J. Gonzales}
\affiliation{Department of Astronomy and Astrophysics, University of California, Santa Cruz; Santa Cruz, CA 95064}

\author{Lea Hirsch}
\affiliation{Astronomy Department, University of California, Berkeley; Berkeley CA, 94720, USA}

\author{Eugene A. Magnier}
\affiliation{Institute for Astronomy, University of Hawaii, 2680 Woodlawn Drive, Honolulu HI 96822}

\author{Sam Ragland}
\affiliation{W. M. Keck Observatory, 65-1120 Mamalahoa Hwy., Kamuela, HI 96743, USA}

\author{R. Michael Rich}
\affiliation{Department of Physics and Astronomy, UCLA, PAB 430 Portola Plaza, Box 951547, LA, CA 90095-1547}

\author{Feige Wang}
\affiliation{Department of Physics, Broida Hall, UC Santa Barbara, Santa Barbara, CA 93106-9530 USA}

\begin{abstract}
We report the discovery of a large ($\sim$8500 km diameter) infrared-bright storm at Neptune's equator in June 2017. We tracked the storm over a period of 7 months with high-cadence infrared snapshot imaging, carried out on 14 nights at the 10 meter Keck II telescope and 17 nights at the Shane 120 inch reflector at Lick Observatory. The cloud feature was larger and more persistent than any equatorial clouds seen before on Neptune, remaining intermittently active from at least 10 June to 31 December 2017. Our Keck and Lick observations were augmented by very high-cadence images from the amateur community, which permitted the determination of accurate drift rates for the cloud feature. Its zonal drift speed was variable from 10 June to at least 25 July, but remained a constant $237.4 \pm 0.2$ m s$^{-1}$ from 30 September until at least 15 November. The pressure of the cloud top was determined from radiative transfer calculations to be 0.3-0.6 bar; this value remained constant over the course of the observations.  Multiple cloud break-up events, in which a bright cloud band wrapped around Neptune's equator, were observed over the course of our observations. No ``dark spot'' vortices were seen near the equator in HST imaging on 6 and 7 October. The size and pressure of the storm are consistent with moist convection or a planetary-scale wave as the energy source of convective upwelling, but more modeling is required to determine the driver of this equatorial disturbance as well as the triggers for and dynamics of the observed cloud break-up events.

\end{abstract}

\section{Introduction} 

The Voyager 2 spacecraft flyby of Neptune in 1989 revealed an extremely dynamic, turbulent atmosphere \citep[][]{smith89, tyler89}. Since then, advances in Earth-based observing, including 10m-class optical/infrared telescopes with adaptive optics systems, the Hubble Space Telescope (HST), the Combined Array for Research in Millimeter-wave Astronomy (CARMA), the Atacama Large (sub-)Millimeter Array (ALMA), and the recently-upgraded Very Large Array (VLA), have permitted multi-wavelength global monitoring of the planet's clouds and deep atmosphere. At infrared wavelengths Neptune shows a striking pattern of bright midlatitude features against a dark background \citep[e.g.][]{roe01, sromovsky01a, max03}. In contrast to images at visible wavelengths, in which Rayleigh scattering produces a relatively uniformly-illuminated planet disk, methane absorption and collision-induced absorption (CIA) by H$_2$ in cloud-free regions makes Neptune's disk dark. In the Kp band (2.2 $\mu$m), the contrast in reflectivity between reflective methane clouds and columns free of discrete upper tropospheric clouds reaches up to two orders of magnitude \citep[e.g., ][]{max03, depater14}.  Infrared observations are useful to probe the conditions under which clouds and storm systems form, the structure and composition of these storms, and their evolution over time.  \citet{sromovsky95, sromovsky01c} and more recently \citet{tollefson18} (along with many other authors) performed optical and infrared cloud tracking to determine Neptune's zonal wind profile. Many authors published infrared spectroscopic data with Gemini \citep[][]{irwin11, irwin14, irwin16} and Keck \citep{max03, depater14, slc16}; these studies included latitudinal mapping of Neptune's methane, characterization of the stratospheric haze layer, and the observation of cloud layers both below and above the tropopause at pressures of 0.3-2 bar and 20-80 mbar, respectively.


An effort to model the circulation on Neptune parallelled these observational studies. Clouds form when humid (rich in condensible species) upwelling air reaches a low enough temperature that condensation occurs, meaning that convective upwelling results in localized cloud systems and downwelling regions tend to remain relatively cloud-free. Combining these physical principles with the observed cloud patterns and deep atmosphere brightness temperature maps led to a hypothesis for Neptune's convection in which air rises from as deep as 40 bars into the stratosphere at midlatitudes, and subsides over the poles and at the equator \citep[][]{depater14}, explaining the cloud bands at Neptune's midlatitudes and the relative paucity of clouds at the equator. 


In this paper we report the discovery of a long-lived cloud complex at Neptune's equator, bright enough in the near-infrared to be observed with even amateur ($\sim$10 inch diameter) telescopes over the second half of 2017. In Section \ref{section_observations} we present observations of the cloud complex with infrared and optical telescopes over roughly seven months from June 2017 to January 2018. We track the position and morphology of the bright cloud feature and perform radiative transfer calculations to estimate the pressure of the cloud top in Section \ref{section_results}. Finally, we explore the implications of these results with respect to the fluid dynamics processes underlying the storm in Section \ref{section_discussion} before summarizing our findings in Section \ref{section_conclusions}.

\section{Observations and Data Reduction}
\label{section_observations} 

We obtained near-infrared and optical images of Neptune over a seven-month period from June 2017 to January 2018 with multiple telescopes; these observations are summarized in Table \ref{professionaldata}, which lists data from large ($>$3 meter) telescopes with adaptive optics (AO) systems as well as HST observations, and Table \ref{amateurdata} (in the Appendix), which lists data from smaller ground-based telescopes that lack AO systems. 


\begin{table}
	\footnotesize
	\begin{tabular}{|c|c|c|c|c|c|}
 & UT Date \& &  & Sub-Observer & Ang. Diam. & \\ 
Telescope & Start Time & Observer & Longitude & (arcsec) & Filters \\ 
\hline 
Keck & 2017-06-26 14:52 & TZ & 298 & 2.31 & H, Kp, CH4S \\ 
Keck & 2017-07-02 12:06 & TZ & 214 & 2.32 & H, Kp, CH4S \\ 
Lick & 2017-07-07 09:47 & Gates & 324 & 2.32 & H, Ks \\ 
Lick & 2017-07-10 10:32 & Gates & 150 & 2.33 & H, Ks  \\ 
Lick & 2017-07-13 10:36 & Lynam/de Rosa & 321 & 2.33 & H, Ks  \\ 
Keck & 2017-07-16 15:08 & Puniwai/TZ & 231 & 2.33 & H, Kp, CH4S \\ 
Keck & 2017-07-24 12:53 & TZ & 152 & 2.34 & H, Kp, CH4S, PaBeta \\ 
Keck & 2017-07-25 15:14 & TZ & 21 & 2.34 & H, Kp, CH4S, PaBeta \\ 
Keck & 2017-08-03 13:25 & Jordan/TZ & 127 & 2.35 & H, Kp, CH4S, PaBeta \\ 
Keck & 2017-08-03 15:26 & Jordan/TZ & 172 & 2.35 & H, Kp \\ 
Lick & 2017-08-06 11:36 & Ammons/Dennison/Lynam & 256 & 2.35 & H, Ks  \\ 
Lick & 2017-08-08 12:00 & Rich/Lepine/Gates & 258 & 2.35 & H, Ks  \\ 
Keck & 2017-08-25 11:23 & Sromovsky/Fry/TZ & 2 & 2.36 & H, Kp \\ 
Keck & 2017-08-26 10:31 & Sromovsky/Fry/TZ & 159 & 2.36 & H, Kp \\ 
Lick & 2017-08-31 08:11 & Crossfield/Gonzales/Gates & 268 & 2.36 & H, Ks  \\ 
Lick & 2017-09-01 10:42 & Crossfield/Gonzales/Gates & 141 & 2.36 & H, Ks  \\ 
Keck & 2017-09-03 10:31 & TZ & 130 & 2.36 & H, Kp, CH4S \\ 
Keck & 2017-09-03 12:57 & TZ & 184 & 2.36 & H, Kp, CH4S \\ 
Keck & 2017-09-04 10:32 & TZ & 306 & 2.36 & H, Kp, CH4S \\ 
Keck & 2017-09-04 12:40 & TZ & 354 & 2.36 & H, Kp, CH4S \\ 
Keck & 2017-09-27 04:56 & Mcllroy/Magnier & 277 & 2.35 & H, Kp, CH4S, PaBeta \\ 
Lick & 2017-10-04 06:47 & Duchene/Oon/Coy/Gates/Lynam & 112 & 2.35 & H, Ks  \\ 
Lick & 2017-10-05 03:52 & Duchene/Oon/Coy/Gates/Lynam & 224 & 2.35 & H, Ks  \\ 
Lick & 2017-10-05 07:29 & Duchene/Oon/Coy/Gates/Lynam & 304 & 2.35 & H, Ks  \\ 
Lick & 2017-10-06 06:08 & Rich/Lepine/Gates & 91 & 2.35 & H, Ks  \\ 
HST  & 2017-10-06 09:02 & OPAL Program & 155 & 2.35 &F467M, F547M, F657M, \\
 & & & & & F619N, F763N, F845M \\
Keck & 2017-10-06 10:54 & Aycock/Ragl & 197 & 2.35 & H, Kp \\ 
HST  & 2017-10-07 02:40 & OPAL Program & 155 & 2.35 &F467M, F547M, F657M, \\
 & & & & & F619N, F763N, F845M \\
Keck & 2017-11-08 04:14 & Alvarez/Licandro & 106 & 2.31 & H, Kp, CH4S \\ 
Lick & 2017-11-29 01:47 & Wang/Gates & 152 & 2.29 & H, Ks  \\ 
Lick & 2017-11-30 02:01 & Melis/Gates & 334 & 2.29 & H, Ks  \\ 
Lick & 2017-12-01 01:46 & Melis/Gates & 145 & 2.28 & H, Ks  \\ 
Lick & 2017-12-02 01:47 & Melis/Gates & 321 & 2.28 & H, Ks  \\ 
Lick & 2017-12-06 02:31 & Hirsch/Gates & 323 & 2.28 & H, Ks  \\ 
Lick & 2017-12-29 02:08 & Melis/Lynam & 48 & 2.25 & H, Ks  \\ 
Lick & 2017-12-31 02:22 & Chen/Lynam & 45 & 2.25 & H, Ks  \\ 
Keck & 2018-01-10 04:38 & Puniwai/McPartland & 58 & 2.24 & H, Kp, CH4S, PaBeta \\     
	\end{tabular}
	\caption{Description of Keck and Lick data used in this publication. ``TZ'' refers to the Keck Twilight Zone observing team---E. Molter, C. Alvarez, I. de Pater, K. de Kleer, and R. Campbell.\label{professionaldata}}
\end{table}

\subsection{Keck and Lick Observations}
\label{keckdata}

\subsubsection{Observing Strategy}

The Keck and Lick data used in this publication were carried out via ``voluntary ToO'' scheduling, which we describe here. We produced an automated script to carry out short (10- to 40-minute) snapshot observations of bright solar system objects at short notice. With these in place, any observer could choose to carry out our observations during their observing time by running the script.  In practice, this occurred mainly during poor weather conditions or twilight hours, when many observations (e.g. spectroscopy of faint targets at optical wavelengths) could not be carried out effectively, and relied heavily on the Observing Assistant (OA) on duty to provide both the impetus and expertise to carry out the observation. The benefits of this model are twofold: telescope time that would have otherwise gone to waste was used for science, and high-cadence short observations were made possible at a classically-scheduled observatory. However, because the observations needed to be short and easy for any classically-scheduled observer at the telescope to carry out, photometric calibration of the data could not be obtained. For this reason, only two of the Keck observations and none of the Lick observations used in this publication have been photometrically calibrated. This observing strategy was pioneered at Keck Observatory as the \textit{Twilight Zone} program.\footnote{https://www2.keck.hawaii.edu/inst/tda/TwilightZone.html}  It had already been in place at Lick Observatory since 2015 and led to one previous publication \citep[][]{hueso17}.

\subsubsection{Data Reduction}

\begin{figure}
	\includegraphics[width = 1.0\textwidth]{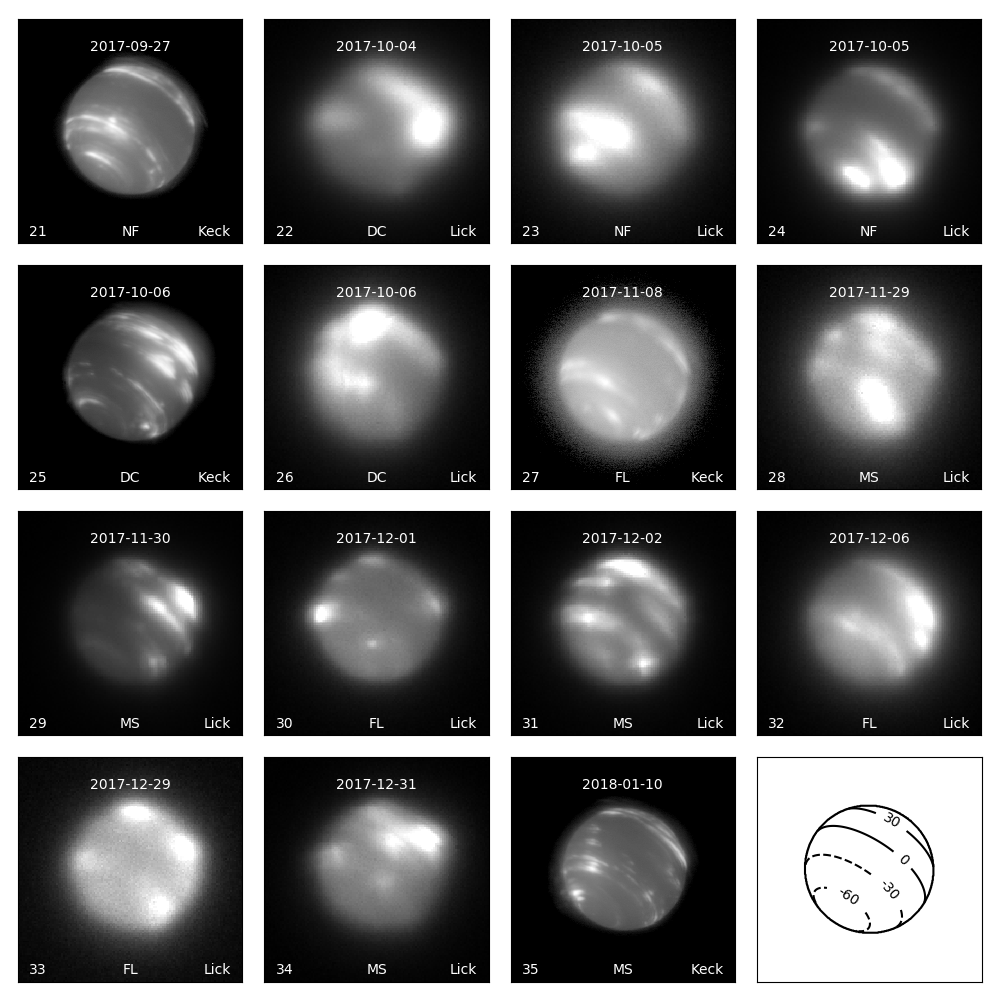}
	\caption{Time series of all H-band Keck and Lick images. The last panel shows the orientation of Neptune relative to the observer. The Keck observations are displayed on a logarithmic scale for better viewing of both bright and faint features. One or two bright equatorial storm features are visible in Panels 1-5, 8, 22, 25, and 26 (labeled DC for Discrete Cloud). Multiple equatorial features or bands are visible in Panels 7, 9-14, 17-20, 28, 29, 31, and 34 (labeled MS for Multiple Spots). One or more faint features are visible near the limb of the planet in panels 27, 30, 32, and 33 (labeled FL for Faint Limb).  No clear equatorial features are visible in Panels 6, 15, 16, 21, 23, or 24 (labeled NF for No Featues).\label{data_thumbnails}}
\end{figure}

Data were obtained on 14 nights from the Keck II telescope on Maunakea, Hawaii. We used the NIRC2 near-infrared camera coupled with the adaptive optics (AO) system, using Neptune itself as the AO guide star. Using the narrow camera on NIRC2, the instrument's smallest pixel scale, yielded a pixel scale of 9.94 mas px$^{-1}$ \citep[][]{depater06}, or $\sim$210 km px$^{-1}$ at Neptune's distance. Data were obtained in the broadband H and Kp filters on all 14 nights, and narrow-band observations in the CH4S and PaBeta filters were obtained when time permitted (see Table \ref{professionaldata}). The images are shown in Figure \ref{data_thumbnails}, and characteristics of each filter are listed in Table \ref{filtertable}.

\begin{figure}
	\includegraphics[width = 1.0\textwidth]{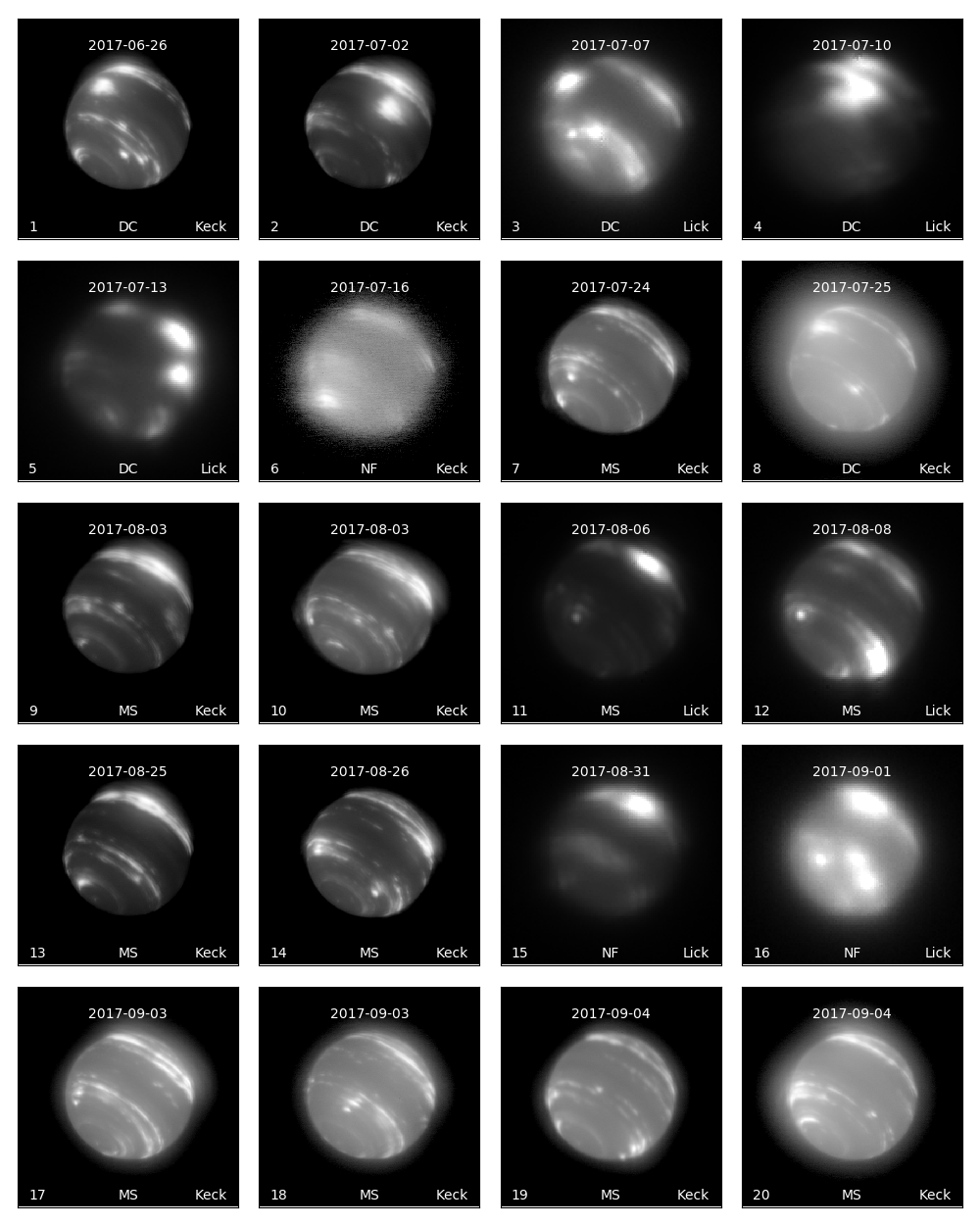}
\end{figure}

Wide-band images were processed using standard data reduction techniques of sky subtraction, flat fielding, and median-value masking to remove bad pixels. Each image was corrected for the geometric distortion of the NIRC2 detector array according to the solution provided by \citet{service16}. Cosmic rays were removed using the \texttt{astroscrappy} package\footnote{https://github.com/astropy/astroscrappy} \citep[affiliated with the community-sourced \texttt{astropy} Python suite;][]{astropy18}. This package implements a version of the standard L.A.Cosmic algorithm \citep[][]{vandokkum01}, which relies on Laplacian edge detection to differentiate cosmic rays from PSF-convolved sources.

Photometric calibration was carried out on 26 June and 25 July using the photometric standard star HD1160 on both dates. This star appears in the UKIRT MKO Photometric Standards list\footnote{http://www.gemini.edu/sciops/instruments/nearir-resources/photometric-standards/ukirt-standards}; its spectral type is A0V and it has J, H, and K band magnitudes of 6.983, 7.013, and 7.040, respectively \citep[in the 2MASS system;][]{cutri03}. Due to our reliance on donated observing time, these were the only dates for which photometry could be obtained. We converted the observed flux densities to units of I/F, the ratio of the observed radiance to that from a normally-illuminated white Lambertian reflector at the same distance from the sun as the target \citep{hammel89}:
\begin{equation}
	\label{if}
	\frac{I}{F} = \frac{r^2}{\Omega} \frac{F_N}{F_\odot}
\end{equation}
where $r$ is Neptune's heliocentric distance in AU, $\pi F_\odot$ is the Sun's flux density at Earth, $F_N$ is the observed flux density of Neptune, and $\Omega$ is the solid angle subtended by one detector pixel. The solar flux density was determined by convolving a high-resolution spectrum from \citet{gueymard04} with the NIRC2 filter passbands. Uncertainties in I/F were set to be 20\% to account for errors in photometry, which we estimated by looking at the difference in flux between the three exposures taken on the standard star in each filter; this 20\% uncertainty is consistent with \citet{depater14}.

To ensure the photometric calibration in H and Kp band from HD1160 was reasonable, we used it to determine the geometric albedo of Neptune's moon Proteus. Proteus was inside the field-of-view of the narrow camera in only one image of the three-point dither on both 26 June and 25 July. The technique we employed to determine the albedo was very similar to that used by \citet{gibbard05} and is summarized in Appendix \ref{appendix_proteus}; the results of that calculation are given in Table \ref{moonphot}. The geometric albedos we found were somewhat higher than the K-band value of $0.058 \pm 0.016$ reported by \citet{roddier97} but in good agreement with \citet{dumas03}, who obtained $0.084 \pm 0.002$ in the HST F160W filter at 1.6 $\mu$m and $0.075 \pm 0.010$ in the HST F204M filter at 2.04 $\mu$m.


\begin{table}
	\begin{tabular}{|c|c|c|c|c|c|}
	Date & Band & $F_{0.2}/F_{tot}$ & Albedo & Error (\%) & Phase Angle ($^\circ$)\\ 
	\hline
	2017-06-26 & H & 0.63 & 0.080 & 21 & 1.8\\
	           & Kp & 0.74 & 0.092 & 21 & 1.8\\
	2017-07-25 & H & 0.30 & 0.073 & 22 & 1.2\\
	           & Kp & 0.29 & 0.100 & 33 & 1.2\\
	\end{tabular}
	\caption{Photometry of Neptune's moon Proteus, used to validate our standard star photometric calibration. The stated error combines the $\sim$20\% photometry error with the estimated additional error from flux bootstrapping (see Appendix \ref{appendix_proteus}).\label{moonphot}}
\end{table}

Data were obtained on 17 nights from the Shane 120-inch reflecting telescope at the UCO Lick Observatory on Mount Hamilton, California. We used the Shane AO infraRed Camera-Spectrograph (ShARCS) camera, a Teledyne HAWAII-2RG detector, coupled with the ShaneAO system and using Neptune itself as a guide star. The pixel scale of the ShARCS images was 33 mas px$^{-1}$, or $\sim$700 km px$^{-1}$ at Neptune. Data were obtained in the broadband H and Ks filters on all 17 nights. Data reduction was carried out using the same procedure as for the Keck data, and the images are shown along with the Keck data in Figure \ref{data_thumbnails}.

\subsubsection{Image Navigation and Orthogonal Projection}
\label{section_nav}

To overlay a latitude-longitude grid onto Neptune and project its ellipsoidal surface onto a map, we followed the same general procedure as used in \citet{sromovsky05}; however, we have recast those codes into the Python programming language and made a few small changes. We summarize the steps here. First, an ellipsoidal model of Neptune was produced with $r_{eq} = 24766$ km and $r_{pol} = 24342$ km as found by Voyager \citep[][]{lindal92}. The model was resized, rotated, and cast into two dimensions to match the angular scale and orientation of Neptune at the time each observation was taken, making use of data from JPL Horizons\footnote{https://ssd.jpl.nasa.gov/horizons.cgi}. Second, the model Neptune was overlain onto the image data and shifted to the location of Neptune in the image.  To achieve this, we employed the Canny edge detection algorithm \citep[implemented by the \texttt{scikit-image} Python package;][]{skimage14}\footnote{skimage.feature.canny; https://scikit-image.org/} to find the edges of Neptune and then simply matched these edges to the edges of the model. The navigation error using this method was the combination of the uncertainty in the shift required to match the model and data (implemented by the \texttt{image\_registration} package\footnote{ image\_registration.chi2\_shifts.chi2\_shift; https://github.com/keflavich/image\_registration}) and the spread in the location of the edges the algorithm found as its parameters were varied over a reasonable range. We found the combined error to be $<0.1$ pixels in all images with acceptable seeing conditions. Third, we interpolated this mapping between image (x, y) coordinates and physical (latitude, longitude) coordinates onto a regular latitude-longitude grid using a cubic spline interpolation \citep[implemented by the \texttt{scipy} Python package;][]{scipy18}\footnote{scipy.interpolate.griddata; https://www.scipy.org/}. Planetographic latitudes were used here and throughout this paper. We chose the grid spacing such that one pixel in latitude-longitude space was the same size as one pixel in image space at an emission angle of zero; that is, the latitude-longitude map was oversampled compared to the data away from the sub-observer point.

The code used for NIRC2 data reduction, navigation, and projection was implemented in Python and has been made publicly available on GitHub\footnote{https://github.com/emolter/nirc2\_reduce}.

\subsection{Observations with non-AO Telescopes}

We alerted the amateur community to the presence of the bright storm feature after it was imaged with Keck on 26 June. A total of 62 near-infrared amateur observations of the feature were made on 33 different nights. Amateur observers D. Milika \& P. Nicholas in fact made the first observation of the storm on 10 June, though it was not recognized as noteworthy until it was later observed with Keck. The bright equatorial feature was also observed with the PlanetCam instrument \citep[][]{mendikoa16} on the 2.2m telescope at Calar Alto Observatory on 11 July. Table \ref{amateurdata} (in the Appendix) summarizes the dates and characteristics of these PlanetCam and amateur observations, and sample amateur images are shown in Figure \ref{amateur_appendix} (in the Appendix).  Images were navigated in WinJupos\footnote{http://jupos.org/gh/download.htm} using the position of Triton as a tie-point for the Neptune latitude-longitude grid; see \citet{hueso17} for a more complete description of this technique.



\subsection{HST Observations}

\begin{figure}
	\includegraphics[width = 1.0\textwidth]{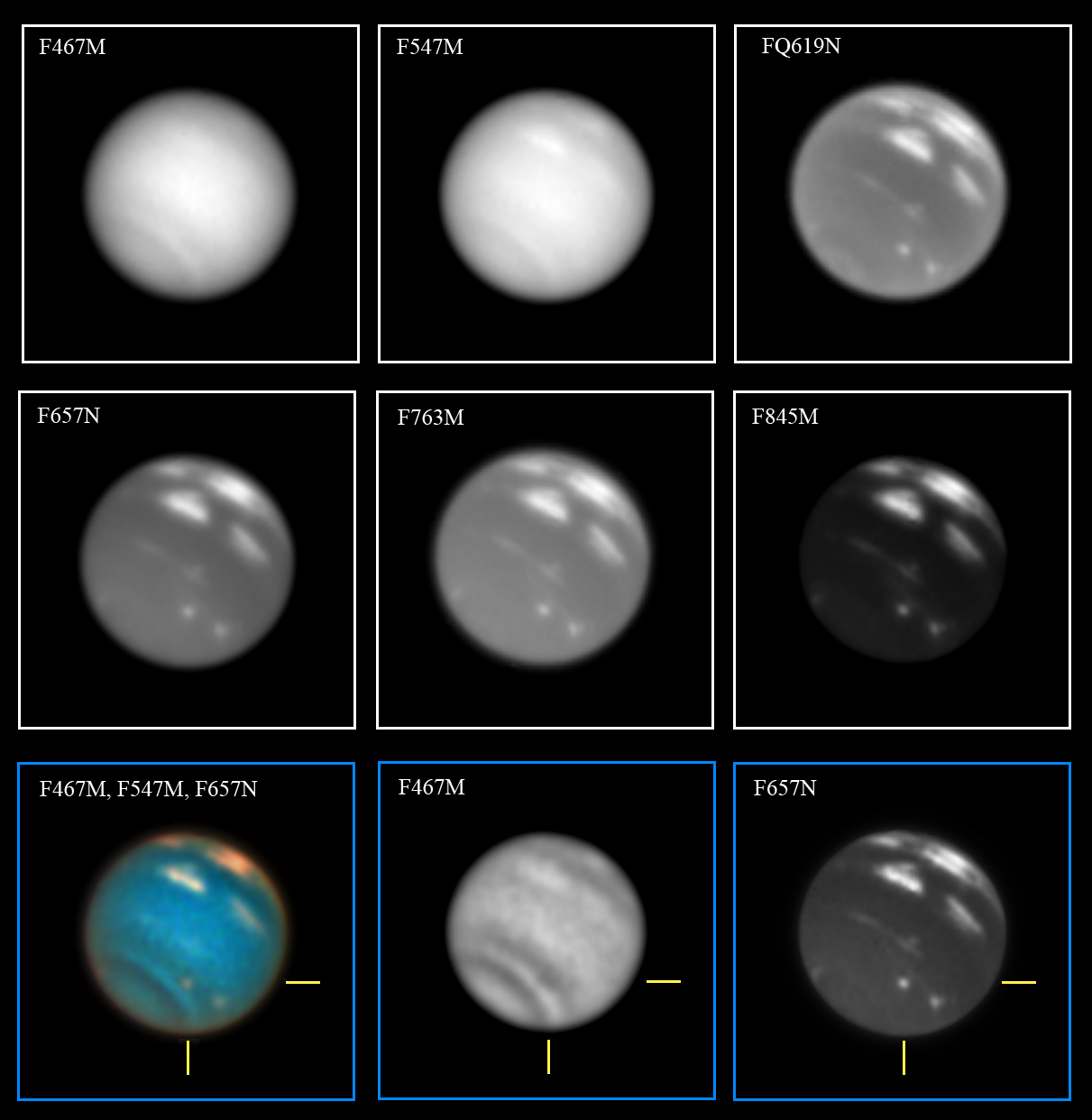}
	\caption{HST images of the equatorial features obtained on 6 October 2017. The two upper rows show images acquired from blue to near-infrared wavelengths, and the lower row shows high-pass versions of images at selected wavelengths. The SDS-2015 dark vortex and its associated bright cloud in red and near IR wavelengths appears highlighted with yellow lines in the color composite and blue image. The bright equatorial clouds do not show any similar dark feature.\label{hstfig}}
\end{figure}

The Hubble Space Telescope (HST) observed Neptune in Cycle 24 on 6 October as part of the Outer Planets Atmospheres Legacy (OPAL) program \citep[][]{simon15b}\footnote{https://archive.stsci.edu/prepds/opal/}. The observations were described by \citet{wong18}, who presented a multi-year study of the southern hemisphere dark vortex SDS-2015.  We processed the images following similar procedures to those described in \citet{wong18}; the images are shown in Figure \ref{hstfig}.

\section{Results}
\label{section_results}

\subsection{Morphological Evolution of the Storm}
\label{section_morph}

\begin{figure}
	\includegraphics[width = 1.0\textwidth]{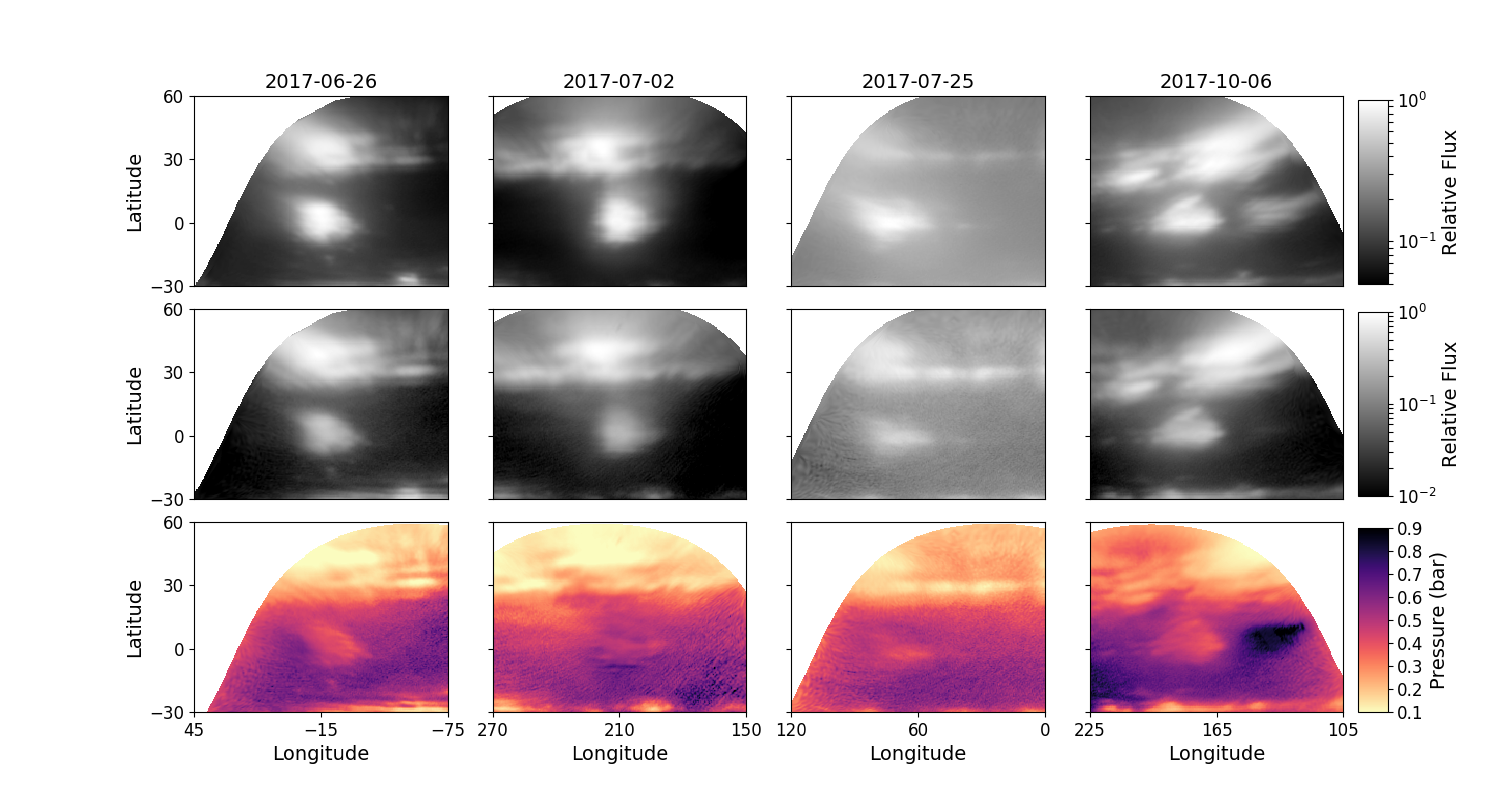}
	\caption{Orthogonally-projected Keck images of the equatorial and northern cloud complexes in H band (\textbf{Top Row}) and Kp band (\textbf{Middle Row}). Images are displayed on logarithmic scales for better viewing of both bright and faint features. Note that the background on 25 July appeared brighter due to poorer atmospheric seeing on that date. \textbf{Bottom Row:} Map of best-fit cloud pressures based on Kp/H ratio in each pixel. These pressures were derived from a radiative transfer model assuming a discrete optically thick cloud (see Section \ref{section_rt}) and are therefore only valid in locations where clouds were visible in H band (top row).\label{storm_morph}}
\end{figure}

An $\sim$8500 km diameter infrared-bright cloud complex was observed at Neptune's equator on several nights from June to December 2017 (see Figure \ref{data_thumbnails}). From at least 26 June to 25 July, this bright equatorial storm remained a single discrete feature, although on 25 July the cloud had elongated compared to 26 June and 2 July and had taken on a somewhat patchy appearance (Figure \ref{storm_morph}). None of the Keck or Lick observations from 3 August to 27 September (13 images on 10 dates) observed a large equatorial storm, but multiple small features at different longitudes were observed at the equator in many of these images. In the first Keck image on 3 August two relatively faint cloud complexes were seen: one thin band near the sub-observer point and another larger group of clouds on the eastern limb of the planet spanning $\sim$15$^{\circ}$ in both latitude and longitude, which may have been remnants of the storm. On 25 and 26 August as well as 3 and 4 September, many small, faint features were observed at various longitudes across Neptune's equator, possibly indicating that the storm sheared apart into an equatorial cloud band. The relative paucity of observations between 25 July and 4 October and the changing drift rate of the storm from 2 June to 25 July (see Section \ref{section_tracking}) made it difficult to determine precisely when the discrete equatorial cloud feature dissipated, since in a single snapshot the storm may have simply been hidden from view on the far side of the planet. However, if the storm maintained its drift rate of $\sim$202 m s$^{-1}$ (see Section \ref{section_tracking}) we should have detected it with Keck on 26 August. We achieved complete longitude coverage on 25 and 26 August and again on 3 and 4 September with Keck, determining with certainty that the storm was not present on Neptune on those dates for any reasonable drift rate. Lick observations on 4 October revealed a bright discrete cloud feature again, and Keck imaging on 6 October captured this feature as well as a detached fainter equatorial cloud roughly 40$^\circ$ east of the main storm.  In all observations in which the equatorial cloud complex was detected, it was coincident in longitude with bright cloud features at the northern midlatitudes from $\sim$30$^\circ$ to $\sim$50$^\circ$. Multiple spots and bands were visible at the equator in 7 Lick observations and one Keck observation from 29 November 2017 to 10 January 2018, revealing that cloud activity on Neptune's equator remained heightened for several months after the reappearance of a large discrete cloud complex. However, individual features could not be tracked over this time period due to the sparse temporal coverage of the data.

\begin{figure}
	\includegraphics[width = 1.0\textwidth]{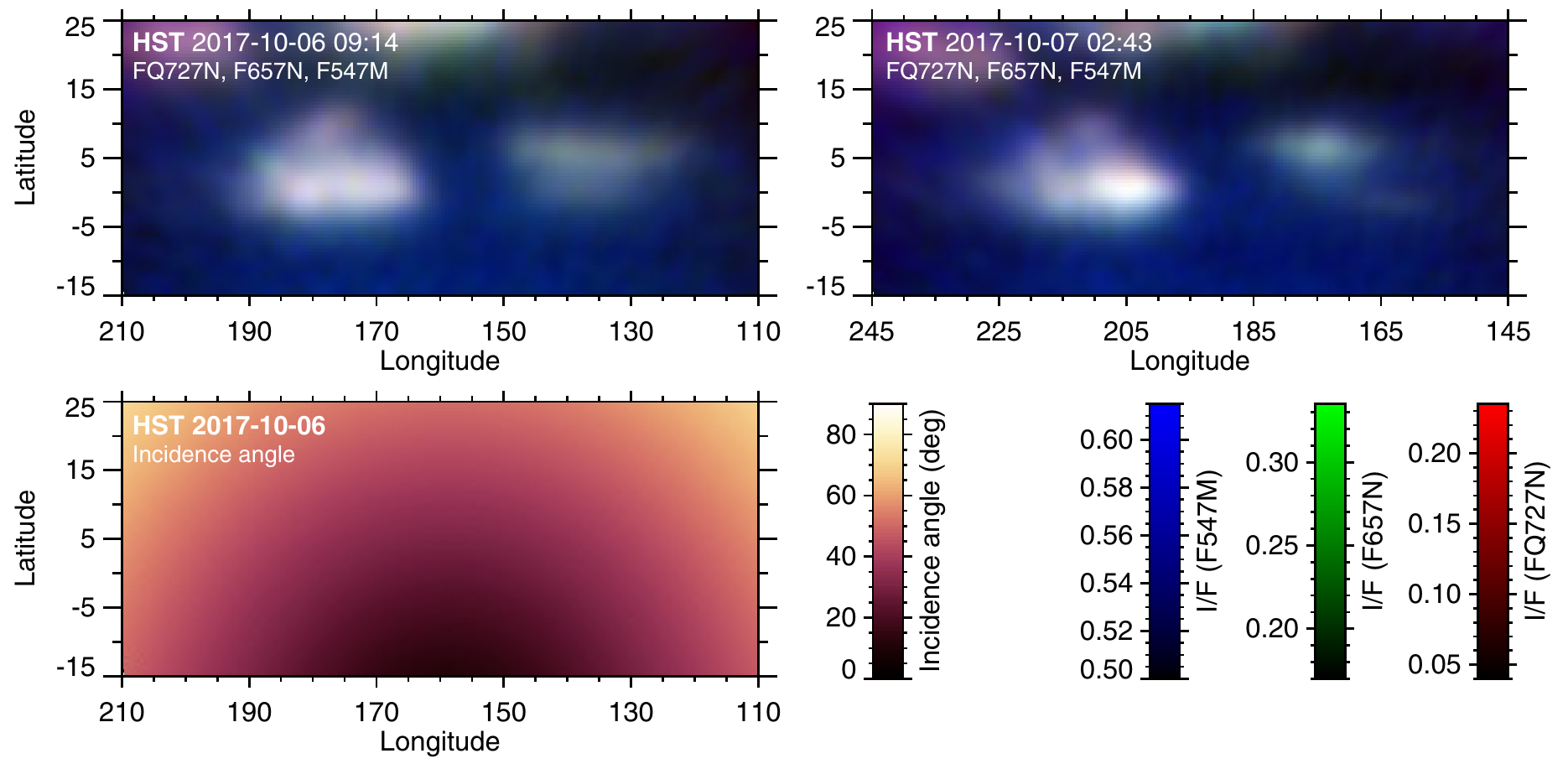}
	\caption{Orthogonally-projected color-composite HST OPAL images from 6 and 7 October. The images reveal changes in storm morphology on $\lesssim$1 day timescales. The lower left panel shows the angle of incidence with respect to the observer.\label{hst_shorttime}}
\end{figure}

The HST observations on 6 October revealed the two bright equatorial clouds observed by Keck faintly in the F467M filter and at progressively higher contrast at increasing wavelengths; the highest contrast was achieved in the F845M filter where methane absorption is most important (see Figure \ref{hstfig}). The bright equatorial storms displayed a similar morphology in the Keck H-band observations on the same date (see Figure \ref{data_thumbnails}, panel 25). Comparing the color-composite images of the equatorial storm taken on 6 October and 7 October reveals that the morphology of both the main storm and its fainter companion cloud varied significantly on timescales of $\lesssim$1 day (see Figure \ref{hst_shorttime}). The dark vortex SDS-2015 was observed near $\sim$45$^{\circ}$S at blue wavelengths \citep[see also][]{wong18} and its companion clouds were visible at red wavelengths, but no dark spot was observed in association with the equatorial storm nor anywhere else on Neptune. Discovery of an equatorial dark spot would have been highly surprising, since \citet{lebeau98} found that anticyclones could not survive within $15^\circ$ of the equator. Their simulations predicted that vortices drifting into the equatorial region would produce extensive perturbations on a global scale, lasting for weeks. This phenomenon was observed on Uranus when the bright ``Berg'' storm drifted toward the equator and dissipated \citep[][]{depater11}. Our observations of persistent equatorial features over several
months is therefore also inconsistent with the dissipation of a vortex that drifted too close to the equator. However, a dark spot obscured by the equatorial cloud complex could not be definitively ruled out by our observations.

\subsection{Feature size determination}
\label{section_size}


To determine the physical extent of the equatorial storm, a 50\% contour was lain down around the bright storm feature in the projected (onto a latitude-longitude grid) images. The longest continuous line segments in the x- and y- directions that fit inside the contour were taken to be the full widths at half-maximum of the storm feature in the zonal and meridional directions, respectively.
This measurement was repeated in both H and Kp band on all four dates on which Keck observed the storm to be a discrete feature, as well as in the HST F845M filter on 6 October; the results are shown in Table \ref{sizes}.

We took the error on these measurements to be one detector pixel. The size of a detector pixel at zero emission angle was $\sim$210 km in the Keck images, so at any other emission angle $\mu = \cos \theta$ the distance on the planet subtended by one pixel was $x \approx (210 / \mu)$ km. However, the most prominent error sources in determining a single value for the cloud's size are the choice of what constitutes part of the cloud, i.e. whether or not the FWHM value is the proper metric, the assumption that the size of the atmospheric disturbance is well represented by the size of the visible region of the cloud, the filter in which the size is measured (which is loosely tied to the pressure level of the cloud), and the short-timescale variability in the cloud's morphology (see Figure \ref{hst_shorttime}).

\begin{table}
	\begin{tabular}{|c|c|c|c|c|c|c|c|}
	     &        & Zonal       & Meridional  &            & Zonal       & Meridional  &                  \\
	Date & Filter & Extent (km) & Extent (km) & Error (km) & Extent ($^\circ$) & Extent ($^\circ$) & Error ($^\circ$) \\ 
	\hline
	2017-06-26 & H & 8315 & 7036 & 345 & 19.2 & 16.6 & 0.8 \\
	           & Kp & 7462 & 6396 & 345 & 17.3 & 15.1 & 0.8 \\
			   \hline
	2017-07-02 & H & 8290 & 8502 & 236 & 19.1 & 20.0 & 0.6 \\
	           & Kp & 7652 & 7439 & 236 & 17.7 & 17.5 & 0.6 \\
			   \hline
	2017-07-25 & H & 15567 & 9045 & 368 & 36.0 & 21.3 & 0.9 \\
	           & Kp & 12411 & 6101 & 368 & 28.7 & 14.4 & 0.9 \\
			   \hline
	2017-10-06 p & H & 12165 & 7341 & 250 & 28.1 & 17.3 & 0.6 \\
	           & Kp & 11117 & 6921 & 250 & 25.7 & 16.3 & 0.6 \\
			   & F845M & 11980 & 7244 & 310 & 27.7 & 17.1 & 0.7 \\
			   \hline
	2017-10-06 s & H & 10907 & 5244 & 442 & 25.2 & 12.3 & 1.0 \\
	           & Kp & - & - & - & - & - & - \\
			   & F845M & 13930 & 6408 & 380 & 32.2 & 15.1 & 0.9 \\
	\end{tabular}
	\caption{Sizes of the equatorial storm on all dates the storm was observed to be a discrete feature by Keck or HST. On 6 October, the ``p'' and ``s'' refer to the primary and secondary storm feature, respectively. The secondary feature was not detected in Kp band.\label{sizes}}
\end{table}

\subsection{Feature tracking}
\label{section_tracking}

We determined the speed of the equatorial storm across the planet by tracking its location over multiple observations.  To do so, we needed to find the latitude and longitude of the storm's center in each image. Because the physical extent of the storm was much larger than one ShARCS or NIRC2 detector pixel and the bright cloud feature had an irregular shape, it was not sensible to determine its center using a Gaussian or elliptical top-hat fit.  Instead, we employed a version of the technique used by \citet{martin12}, which proceeded as follows. First, we defined a large box around the entire bright cloud region. Second, we computed contours around the brightest region of the storm at many levels (from 68\% to 95\% in intervals of 0.01\%; the particular choice of starting and ending values and step size had very little effect on the result). Third, we determined the centroid of each contour.  Finally, we took the mean of these centroid positions as the derived feature center, and took the standard deviation in the retrieved centers to be the $1\sigma$ error on that value. The feature tracking error dominated over the error in the planet's location on the detector (see Section \ref{section_nav}). We note that this technique found the brightest region of the storm and was therefore sensitive to changes in the storm's morphology, which occur on short timescales ($\lesssim$1 day; see Figure \ref{hst_shorttime}), and slightly sensitive to the storm's position with respect to the limb due to limb-darkening/brightening effects.


Feature locations in Amateur and PlanetCam images were measured using the WinJupos software, which permits determination of locations of features across the planet. Measurements were obtained by two of us (R.H. and M.D.) and sometimes by the individual observers, and these two or three location determinations were found to be coincident within the estimated uncertainty. The measurement uncertainty was determined by marking the center of the equatorial bright feature bye eye 3-5 times and observing the dispersion in the measurements.

\begin{figure}
	\includegraphics[width = 1.0\textwidth]{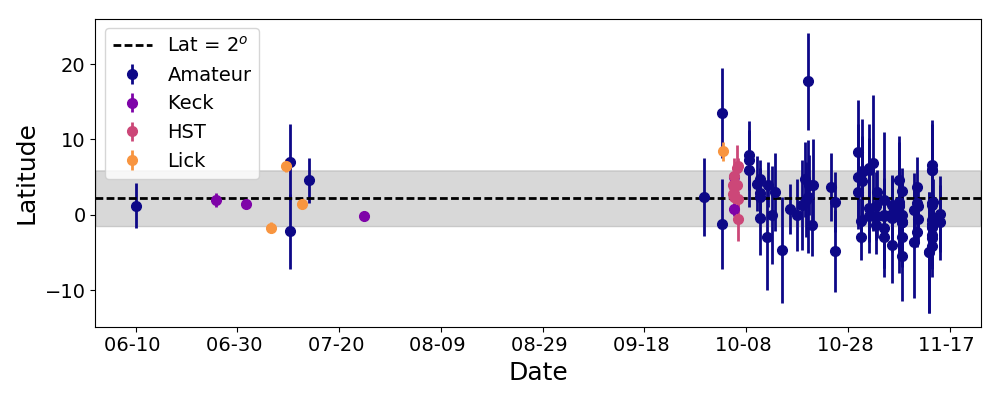}
	\caption{Latitude of the bright equatorial feature over time. The grey region represents the $1\sigma$ dispersion in the latitude measurements.\label{lat_tracking}}
\end{figure}

\begin{figure}
	\includegraphics[width = 1.0\textwidth]{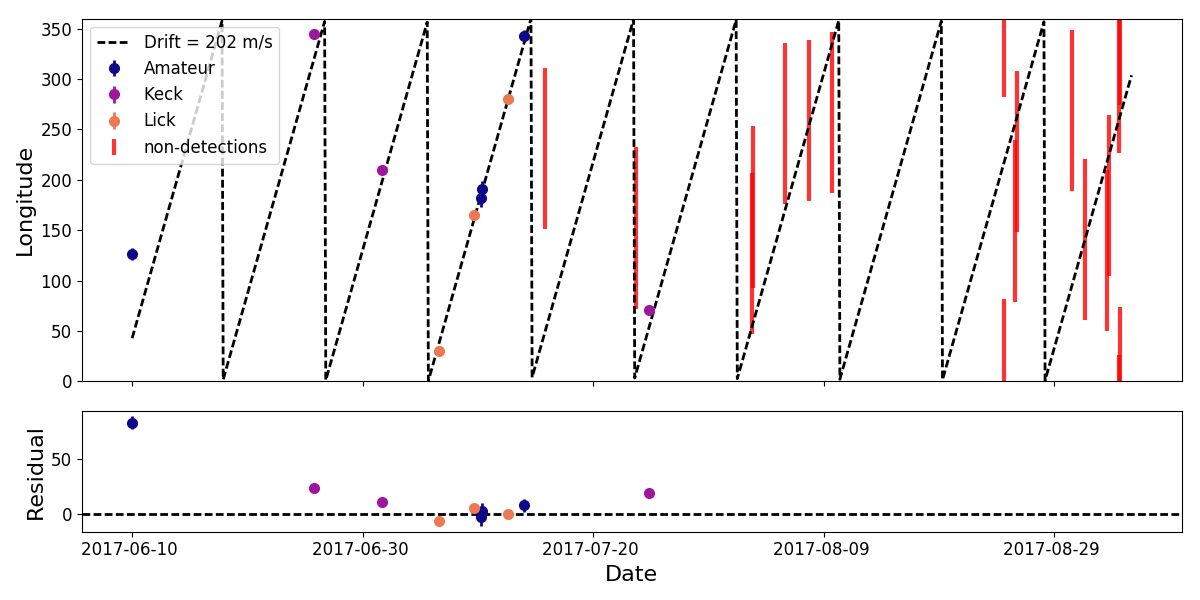}\\
	\includegraphics[width = 1.0\textwidth]{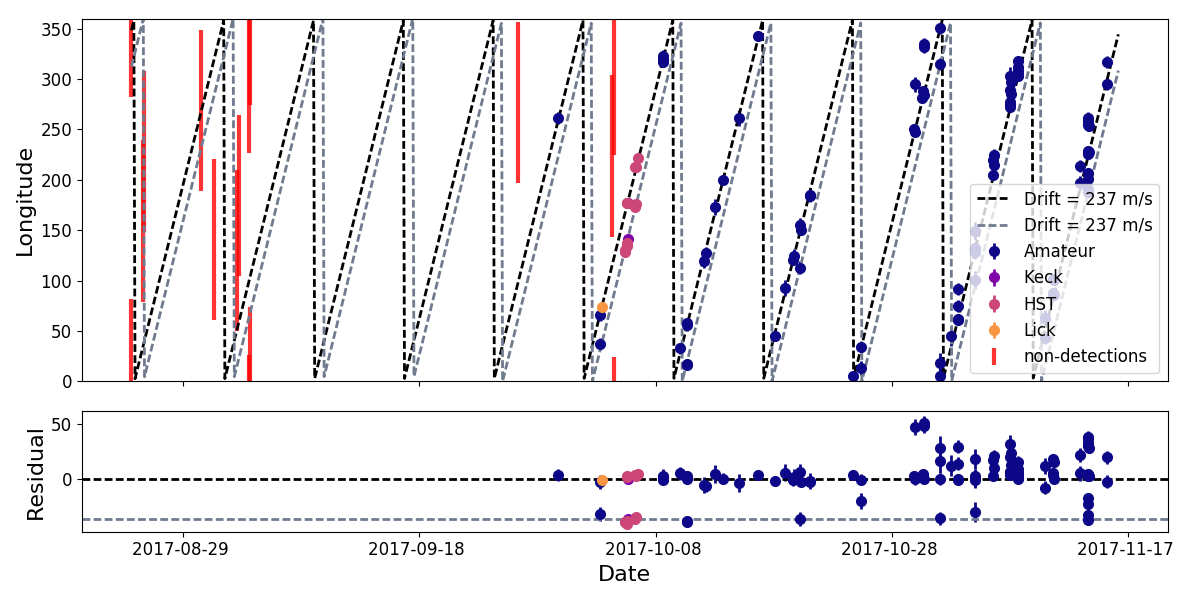}
	\caption{Longitude of the bright equatorial feature over time. The longitude coverage (at the equator) of observations in which a bright equatorial storm was not detected are shown as red bars. Observations are split into two epochs to facilitate visualizing the two different wind speed fits. The black and gray lines are wind speed fits to the main equatorial feature and the detached secondary feature respectively.\label{lon_tracking}}
\end{figure}

The results of our feature tracking are shown in Figures \ref{lat_tracking} and \ref{lon_tracking}. The storm (when present) was stable in latitude, remaining within $\pm$5$^{\circ}$ of the equator over the entire time baseline of our observations. The mean latitude of the feature center was found to be $2.2^{\circ}$ North with a standard deviation of $\pm3.7^{\circ}$.

We derived the longitude drift rate of the storm by fitting our longitude tracking data to a linear model; the details of this process are explained in Appendix \ref{appendix_tracking}.  
Over the first six weeks of observations from 10 June to 25 July, we find a best fit drift rate of 197 $\pm$ 3 m s$^{-1}$. However, this drift rate provides a relatively poor fit to the storm location on 10 June, 26 June, and 25 July: the predicted location of the feature center lies entirely outside the $\sim$20$^\circ$ feature. 
Using instead only the six data points from 2 July to 14 July in the fit, i.e., where the sampling is densest, yields a drift rate of 201.7 $\pm$ 2.2 m s$^{-1}$ and again fails to fit the data points before 2 July or after 14 July (see Figure \ref{lon_tracking}). This implies that the storm's drift rate varied on timescales of a few to tens of days over the first epoch.

Two distinct bright equatorial storms offset by $\sim$50$^{\circ}$ longitude were visible from 28 September to at least 1 November.  The brighter of the two storms is best fit by a constant drift rate of 237.4 $\pm$ 0.20 m s$^{-1}$, and the good fit from 28 September to 1 November implies that a constant drift rate provides a good model for these data. The same drift rate also fits the secondary storm over this epoch, though the data are sparser. On and after 1 November, many equatorial features were observed at different longitudes on the same nights, and the data are not of sufficient resolution or time coverage to track individual features without confusion.


\subsection{Radiative transfer modeling}
\label{section_rt}

\subsubsection{The SUNBEAR Radiative Transfer Code}
\label{subsection_model}

We employed an in-house radiative transfer (RT) code based on the \texttt{disort} module \citep[][]{stamnes88}, a parallelized RT equation solver. The code, which we call SUNBEAR (Spectra from Ultraviolet to Near-infrared with the BErkeley Atmospheric Retrieval), has been adapted to \texttt{Python} based on \texttt{pydisort} \citep[][]{adamkovics16}\footnote{https://github.com/adamkovics/atmosphere/blob/master/atmosphere/rt/pydisort.py}, and used previously for solar system observations on Titan, Uranus, and Neptune \citep[][]{adamkovics16, kdk15, slc16} at infrared wavelengths. SUNBEAR has been extended to visible wavelengths in order to analyze HST data by implementing several scattering processes that are important at visible wavelengths: Rayleigh scattering, Rayleigh polarization, and Raman scattering. Rayleigh scattering was computed by calculating the total Rayleigh-scattering cross section per molecule \citep[][]{mccartney76}, where values for the molecular depolarization and reflective indices came from \citet{allen63}. Rayleigh polarization, which increases the reflectivity of Neptune's atmosphere at high scattering angles and produces an effect as large as 9\% even at zero phase angle \citep[][]{sromovsky05a}, was treated following the empirical approximation developed in \citet{sromovsky05a}. This approximation agrees to the $\lesssim$1\% level for cloud-free or cloud-opaque atmospheres. Raman scattering was taken into account using the semi-empirical approximation discussed in \citet{karkoschka94, karkoschka98}; this technique transforms between the Raman and non-Raman parts of spectra by assuming that the measured geometric albedo is a linear combination of terms involving the spectrum without Raman scattering. \citet{sromovsky05b} found that this approximation is accurate at short wavelengths, but underperforms at longer wavelengths within the methane absorption bands. \citet{karkoschka09} improved their empirical spectral dependencies based on these results, and we used their parameters within SUNBEAR.

\subsubsection{Background Model}
\label{subsection_background}

\begin{figure}
	\includegraphics[width = 0.5\textwidth]{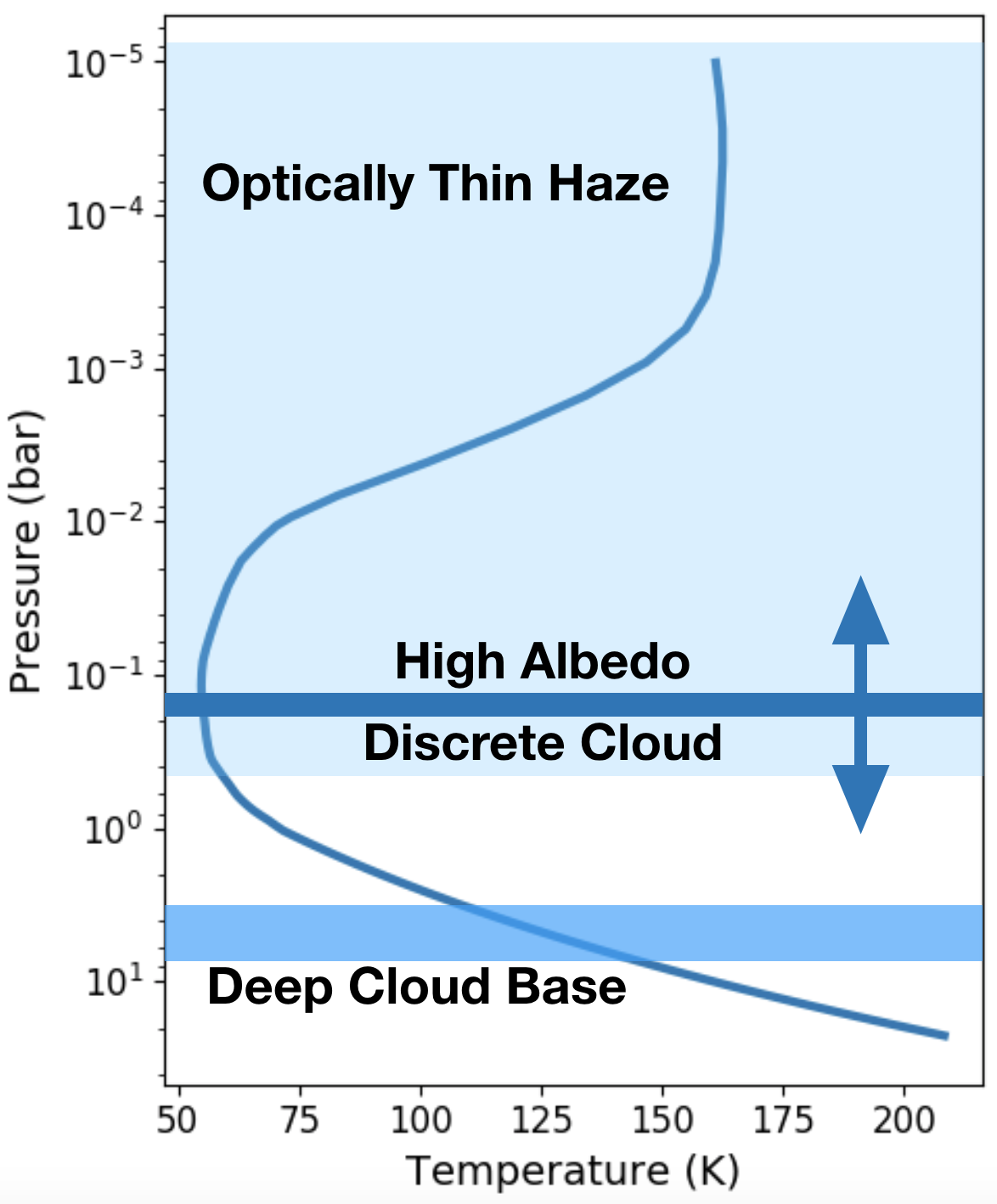}
	\includegraphics[width = 0.5\textwidth]{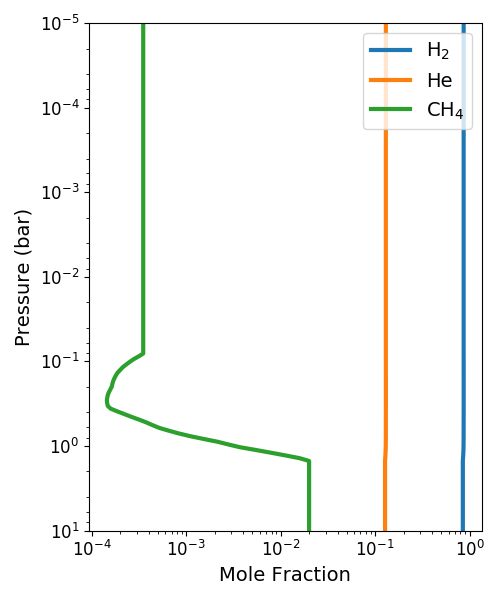}
	\caption{\textbf{Left:} The temperature-pressure profile of Neptune's atmosphere used in our radiative transfer model. The cloud layers in our radiative transfer model are overlain in blue; the arrows on the high-albedo discrete cloud indicate that we changed the pressure of this layer to fit our observations. \textbf{Right:} Vertical abundance profiles of gases in our model.\label{rt_model}}
\end{figure}

We input a temperature-pressure profile and gas abundance profiles appropriate for Neptune's atmosphere \citep[][]{depater14, slc16}, along with an optically thin haze at pressures less than 0.6 bar and an optically thick cloud layer at 3.3 bar; see Figure \ref{rt_model}. This ``background" model was identical to the best-fit model of \citet[][]{slc16} (labeled \textit{2L\_DISORT} in that paper) retrieved from Keck OSIRIS spectral data of Neptune's dark regions, but with one modification. To account for the increase in haze albedo at visible wavelengths inferred from HST spectroscopic data \citep[e.g.,][]{karkoschka11}, the single-scattering albedo $\varpi$ of the optically thick cloud at 3.3 bar was allowed to smoothly vary from 0.45 longward of 1.6 microns, consistent with \citet{slc16}, to 0.99 shortward of 0.5 microns, consistent with \citet{karkoschka11}, according to the following equation: 
\begin{equation}
	\varpi = 0.45 + (0.99 - 0.45)\Big[1 + e^{\frac{\lambda - \lambda_t}{0.1}}\Big]
\end{equation}
The transition wavelength $\lambda_t$ between the two $\varpi$ regimes was set to be 0.8 $\mu$m, and the 0.1 in the denominator of the exponential, which sets the ``sharpness'' of the transition between the two albedo values, was also set arbitrarily to a qualitatively reasonable value. Note that the albedo correction at UV wavelengths from \citet{karkoschka11} is not important longward of 0.4 $\mu$m, so it was ignored here.

\begin{figure}
	\includegraphics[width = 1.0\textwidth]{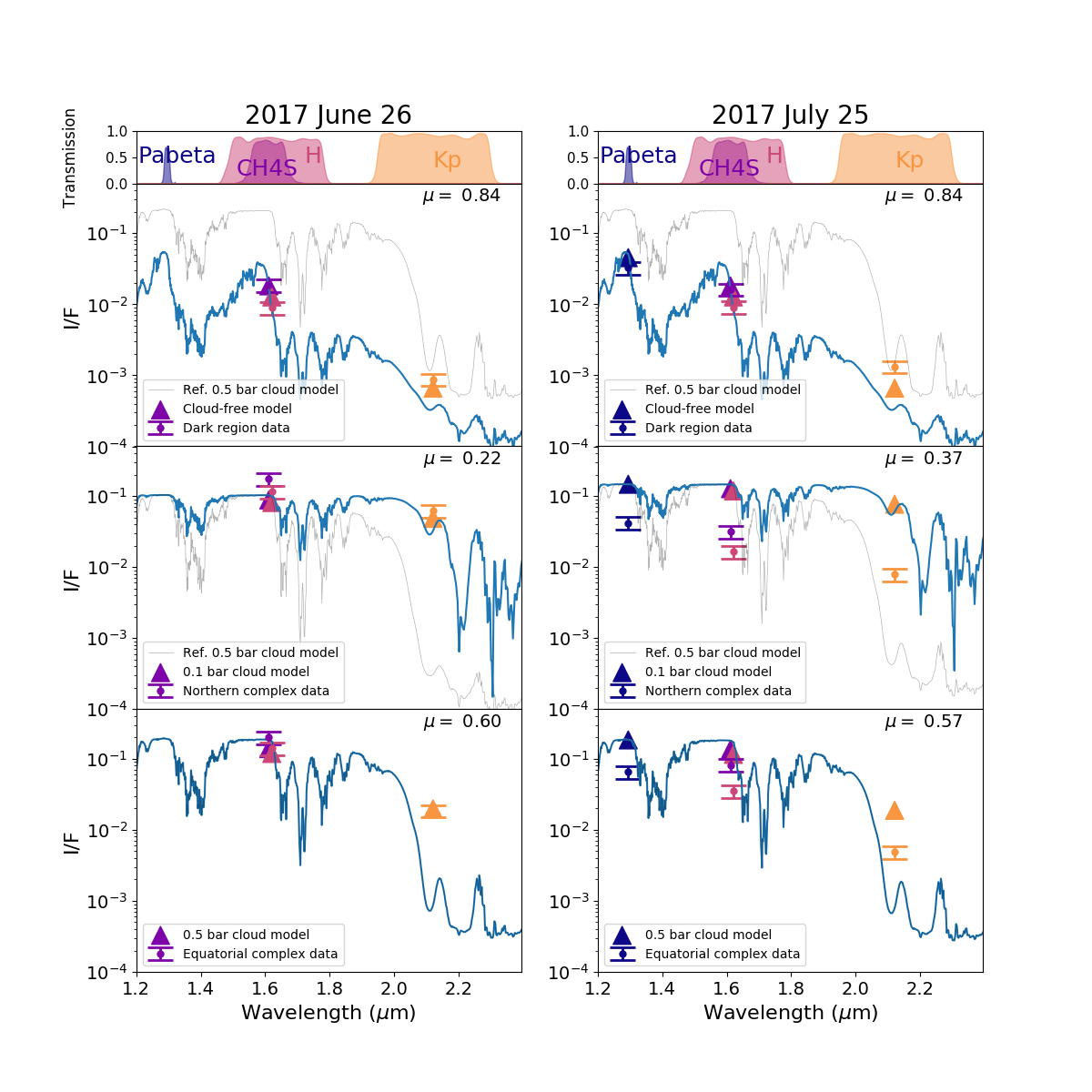}
	\caption{Model I/F values compared to data from 26 June (\textbf{left}) and 25 July (\textbf{right}) for a background region free of discrete upper tropospheric clouds (\textbf{top}), the northern cloud complex (\textbf{middle}), and the equatorial storm (\textbf{bottom}). Triangles represent the model values in each NIRC2 band, derived by convolving the model spectrum (shown here in blue) with the filter passbands. Models in all plots used the reference cloud parameters, varying only pressure and changing $\mu$ to the appropriate value for that cloud. The thin gray line in each panel shows a model spectrum generated using the reference parameters and a discrete cloud pressure of 0.5 bar (identical to the bottom panels, but using the appropriate value of $\mu$) to facilitate visualization of differences between the models.\label{rt_data}}
\end{figure}

We confirmed that the ``background'' model fit our NIRC2 data in regions where discrete upper tropospheric clouds were not observed by convolving the model spectrum from the RT code with the NIRC2 filter bandpasses\footnote{https://www2.keck.hawaii.edu/inst/nirc2/filters.html}. The results of this exercise for our 26 June and 25 July data are shown in Figure \ref{rt_data}. The model fit within the error bars of the data for all filters except Kp on 25 July. We do not view this as a severe problem for the model because the absolute Kp band I/F values in Neptune's dark regions were so minuscule that small systematic modeling errors may have led to large relative offsets. For example, \citet{irwin16} noticed short timescale variablity in the K band in their VLT SINFONI spectrograph observations, which they attributed to changes in the single-scattering albedo. Scattered light from the bright storms on other regions of Neptune may have also contributed to an observed brightening compared to the model, especially in relatively poor seeing as on 25 July; however, since we are interested in the bright cloud regions themselves, scattered light systematics are not a serious concern. 

\begin{figure}
	\includegraphics[width = 1.0\textwidth]{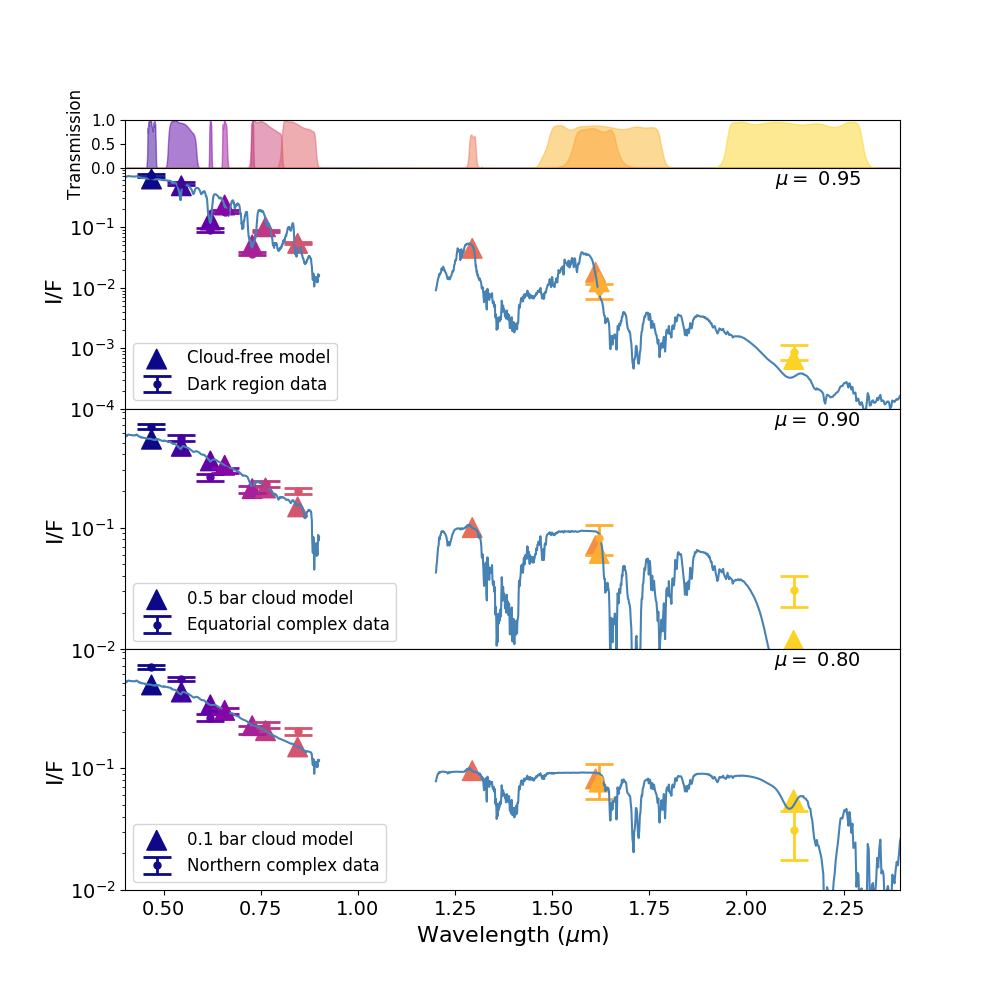}
	\caption{Model I/F values compared to combined Keck and HST data from 6 October for a background region free of discrete upper tropospheric clouds (\textbf{top}), the northern cloud complex (\textbf{middle}), and the equatorial storm (\textbf{bottom}). Triangles represent the model values in each filter, derived by convolving the model spectrum (shown here in blue) with the filter passbands. Models in all plots used the reference cloud parameters, varying only pressure and changing $\mu$ to the appropriate value for that cloud.\label{rt_hstkeck}}
\end{figure}

HST OPAL data and Keck data were taken within two hours of each other on 6 October; however, the Keck data were not photometrically calibrated, so in order to model both datasets together it was necessary to ``bootstrap'' a photometric calibration to the Keck data. To do so, we assumed that the brightness of the dark regions in the two Keck filters was identical (at a given emission angle) on 26 June and 6 October, and then scaled the photon counts in the 6 October images accordingly. This assumption is reasonable because the same background model fit the Keck data on 26 June and 25 July within our uncertainty. Also, the timescale of H-band variability in the haze has been observed to be much longer than the $\sim$4 months between 26 June and 6 October \citep[][]{hammel07, karkoschka11b}. The comparison between our ``background'' model and the combined HST and Keck data is shown in Figure \ref{rt_hstkeck}.

\subsubsection{Discrete Cloud Model}
\label{subsection_discrete}

We inserted an optically thick discrete cloud layer into the ``background'' model to simulate the equatorial storm.  The cloud layer had the following properties, which we refer to as the ``reference'' model: $\tau = 10.0$ was the optical depth; $h_f = 0.05$ was the fractional scale height; $g = 0.65$ was the Henyey-Greenstein parameter; $r_p = 1.0$ $\mu$m was the peak radius in the \citet{deirmendjian64} haze particle size distribution;
\begin{equation}
	\varpi = 0.9 + (0.99 - 0.9)\Big[1 + e^{\frac{\lambda - \lambda_t}{0.1}}\Big]
\end{equation}
was the single-scattering albedo. Since the optical depth, single-scattering albedo, and phase function were all specified, the particle size distribution $r_p$ was only used to set the wavelength dependence of the scattering cross-section and was not truly an independent parameter. The reference cloud model was based on the properties derived by \citet{irwin11, irwin14}, who used the NEMESIS radiative transfer code to model many Neptune infrared cloud spectra observed by the SINFONI spectrograph on the Very Large Telescope (VLT). Their best-fit model spectra varied rather widely in aerosol parameters, and since an equatorial cloud complex similar to what we observed had never been seen before, a good \textit{a priori} guess at the cloud properties was difficult. Nevertheless, those authors favored moderately forward scattering ($g = 0.6-0.7$) and moderate- to high-albedo ($\varpi = 0.4 - 1.0$) clouds in most cases. We chose to model a very compact cloud layer because this was the simplest possible assumption in absence of constraining data. A short description of, reference model values of, and bibliographic references for all of the discrete cloud parameters are summarized in Table \ref{summarize_params}, and we refer the reader to Appendix A of \citet{slc16} for additional explanation of the way these parameters were implemented in SUNBEAR.

\begin{table}
	\begin{tabular}{|c|c|c|c|c|}
	 & & Reference & & Reference(s) \\
	Parameter & Description & Value & Varied? & \& Notes \\ 
	\hline
	$P_m$ & cloud pressure & - & Yes & - \\
	\hline
	$\tau$ & optical depth & 10.0 & Yes & assumed optically thick \\
	 & & & & unless poor fit \\
	 \hline
	$\varpi_{IR}$ & single-scattering albedo & 0.9 & No & \citet{irwin11, irwin14} \\
	 & at IR wavelengths & & & \\
	 \hline
 	$\varpi_{vis}$ & single-scattering albedo & 0.99 & No & \citet{karkoschka11} \\
 	 & at visible wavelengths & & & \\
	 \hline
	$g$ & Henyey-Greenstein parameter & 0.65 & No & \citet{irwin11, irwin14} \\
	\hline
	$r_p$ & peak particle radius & 1.0 $\mu$m & No & \citet{deirmendjian64} \\
	\hline
	$h_f$ & fractional scale height & 0.05 & No & assume very compact \\
	\end{tabular}
	\caption{Summary of discrete cloud model parameters used in this paper. See \citet{slc16} for more complete descriptions of the meaning of these parameters.\label{summarize_params}}
\end{table}



\begin{figure}
	\includegraphics[width = 0.5\textwidth]{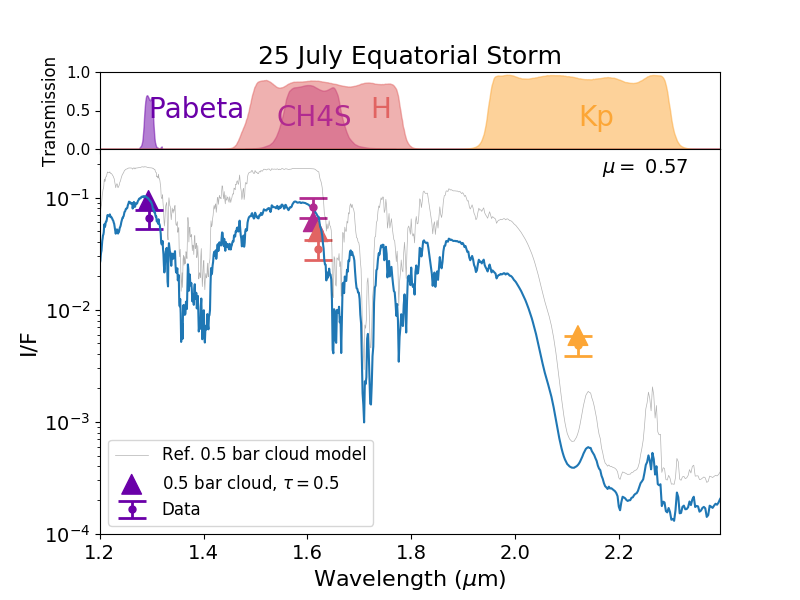}
	\includegraphics[width = 0.5\textwidth]{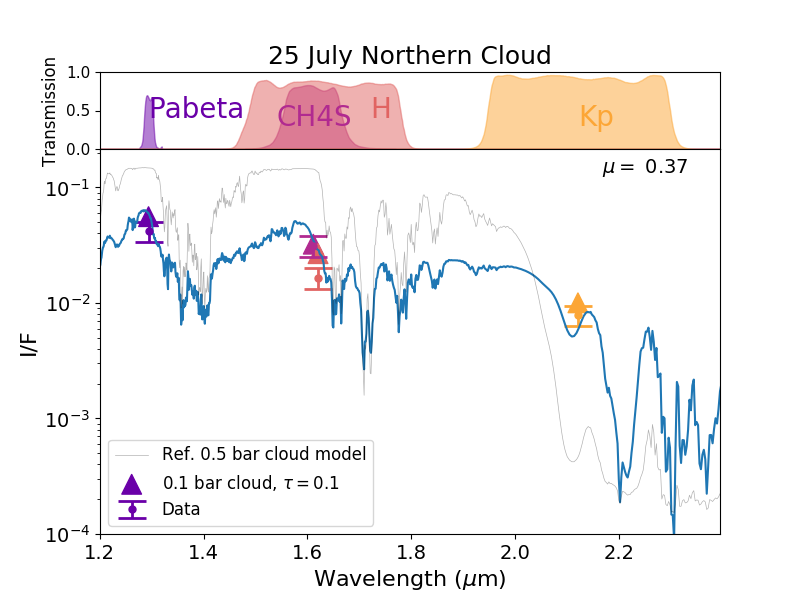}
	\caption{Models of the northern cloud complex and equatorial on 25 July. The models were the same as the reference model (see Figure \ref{rt_data}) but with opacities $\tau = 0.5$ for the equatorial storm and $\tau = 0.1$ for the northern complex. The thin gray line in each panel shows a model spectrum generated using the reference parameters and a discrete cloud pressure of 0.5 bar (identical to the bottom panels in Figure \ref{rt_data}, but using the appropriate value of $\mu$) to facilitate visualization of differences between the models.\label{bestfit_jul25}}
\end{figure}

The pressure $P_m$ of this cloud was varied in steps of $\log P_m = 0.25$ from 10 bar to 0.01 bar to produce a suite of ``discrete cloud'' spectra for clouds from the deep troposphere to the upper stratosphere. We reran these models for values of the emission angle $\mu$ from 0.1 to 1.0 in steps of 0.1, ending up with a grid of spectra in $(P_m, \mu)$ space. We then interpolated across this grid to find the spectrum of any $(P_m, \mu)$ pair. This technique took advantage of the fact that spectra are continuous functions of their labels; that is, changes in cloud pressure, emission angle, or any other input parameter produce smooth changes in the resulting spectrum. This type of interpolation is widely used to fit stellar spectra \citep[e.g., ][]{rix16}.
The model spectra were convolved with the NIRC2 filter bandpasses to produce model reflectivities in each filter for a cloud of a given pressure. Then we determined the average reflectivity of the cloud core in the data by averaging all the pixels in a 90\% contour around the brightest pixel in the cloud, and finally compared the model reflectivities to these data; those fits can be seen in Figure \ref{rt_data}.  On 26 June the reference model provided a good fit to both the equatorial storm and northern complex data for tropospheric cloud layers at $\sim$0.5 bar and $\sim$0.1 bar, respectively; however, on 25 July the reference model fit neither the equatorial storm nor the northern complex for any values of the cloud pressure. We assumed this difference was caused by a decrease in the opacity of the clouds on 25 July compared to 26 June. This interpretation was favored because the equatorial storm appeared to take on a patchy appearance on 25 July, and changes in microphysical cloud parameters ($\varpi$, $g$, $r_p$) for a given cloud type are relatively small on Earth \citep[e.g.][]{baum05}. Good fits to the July 25 data were achieved using optical depths $\tau = 0.1$ for the northern complex and $\tau = 0.5$ for the equatorial storm (see Figure \ref{bestfit_jul25}). We caution that these assumed cloud properties are not a unique fit to the NIRC2 data, as large degeneracies between parameters are present \citep[e.g.,][]{depater14}. It was not possible to retrieve parameters independently, since each of the three haze layers was parameterized by five parameters ($\varpi$, $\tau$, $P_m$, $h_f$, and $g$) but we only had three or four spectral data points on each observation date. The equatorial storm observed on 6 October with combined HST and bootstrapped Keck data was well fit by the reference discrete cloud model at 0.5 bar pressure (see Figure \ref{rt_hstkeck}), pointing to an increase in reflectivity of the storm at visible wavelengths. In our model this increase in reflectivity was achieved via an increased single-scattering albedo; however, this solution is not unique. The particle size, optical depth, and Henyey-Greenstein parameter may also change at visible wavelengths compared to infrared wavelengths, and it is possible to fit the sparse available data using many different combinations of these parameters. In addition, the function we used to smoothly vary the single-scattering albedo was purely empirical, and could also be tuned. All of the models we present here should be taken as only one of many possible physical interpretations of the data.

\begin{figure}
	\includegraphics[width = 0.7\textwidth]{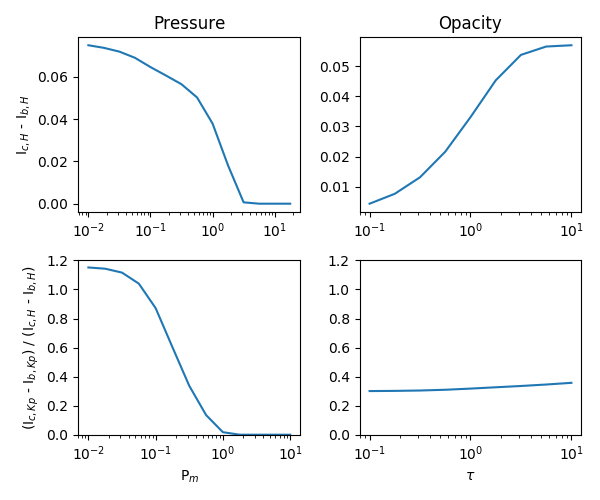}
	\caption{Effect of varying the cloud pressure and opacity in our model on the background-subtracted Kp/H ratio. It can be seen in the bottom row that the ratio depended strongly on the cloud pressure but very little on the opacity. The model had microphysical properties $\varpi = 0.75$, $g = 0.65$, $h_f = 0.05$, and $\mu = 1.0$. \label{new_ratio_plot}}
\end{figure}

The Kp/H ratio in the upper troposphere depends strongly on the cloud pressure $P_m$ and the opacity $\tau$ but only weakly on microphysical cloud properties.
Physically, this is because the pressure of a reflecting cloud layer changes the path length of a photon through the atmosphere before it scatters off the cloud. The path length greatly affects the reflectivity in Kp-band near 2.2 $\mu$m because methane absorption is very strong at those wavelengths, and therefore a photon traveling through more atmosphere has a higher chance of being absorbed. On the contrary, the path length has little effect on the reflectivity in H and CH4S bands near 1.6 $\mu$m because these are much less affected by molecular absorption in the stratosphere and upper troposphere, so the I/F value in those bands is almost entirely governed by scattering off the cloud layer itself. Changing the opacity of the cloud produces a similar effect: a less opaque cloud permits longer path lengths through the atmosphere, and in Kp-band the photons traveling these paths have a high probability of being absorbed whereas in H-band the photons may also backscatter from a haze particle or the deep H$_2$S cloud base.
\citet{depater11} showed that by subtracting the background I/F value from the I/F value of the discrete cloud, the opacity can be eliminated and the ratio equation
\begin{equation}
	\label{kpheq}
	\frac{I_{c,Kp} - I_{b,Kp}}{I_{c,H} - I_{b,H}} = \frac{I_{Kp} (P_m) - I_{b,Kp}}{I_{H} (P_m) - I_{b,H}}
\end{equation}
is obtained, where $I_c$ is the intensity at the location of the discrete cloud, $I_b$ is the intensity of the background, and $I(P_m)$ is the radiance of a very optically thick model discrete cloud at pressure $P_m$ (i.e. our reference model cloud). Equation \ref{kpheq} can be used to place an approximate constraint on $P_m$ independently of opacity by simply finding the pressure at which the left-hand side and right-hand side of the equation are equal; this idea was applied to clouds on Uranus by \citet{depater11} and \citet{sromovsky12}. The solutions to Equation \ref{kpheq} as a function of pressure are shown for our data in Figure \ref{new_ratio_plot}. The figure shows that the background-subtracted Kp/H ratio varies from $0.05-1.1$ from 0.1 to 1 bar, defining the pressure range over which this ratio is a useful pressure probe. This technique can also be used to crudely approximate cloud pressures without absolute flux calibration from photometry, as long as the telluric atmospheric transmission between 1.6 $\mu$m and 2.2 $\mu$m was not strongly wavelength-dependent on the night of the observations.  By evaluating Equation \ref{kpheq} at every pixel in our Keck images, we produced a rough spatial map of the pressures of Neptune's cloud tops on each date for which the equatorial storm was observed with Keck. These maps are shown in Figure \ref{storm_morph}. The validity of this technique was confirmed by fitting models to data for the two photometrically-calibrated Keck datasets. While the overall reflectivity of the clouds changed between 26 June and 25 July, the flux density ratio (inside the 90\% contour) between filters did not change significantly---in the equatorial storm the Kp/H ratio was 0.13 on both dates, and in the northern complex the Kp/H ratio was 0.54 on 26 June and 0.47 on 25 July---and Equation \ref{kpheq} found cloud pressures of 0.5 bar and $\lesssim$0.1 bar for the equatorial and northern clouds, respectively, in agreement with our radiative transfer modeling. It should be noted that the maps in Figure \ref{storm_morph} are only valid at locations where a discrete opaque cloud was detected, since they are based on a radiative transfer model that assumes such a cloud is present. Also, some of the clouds in the images lay at pressures less than 0.1 bar or greater than 1.0 bar; in those cases the pressures determined by this technique should be treated as upper or lower limits, respectively.

\section{Discussion}
\label{section_discussion}

\begin{figure}
	\includegraphics[width = 0.7\textwidth]{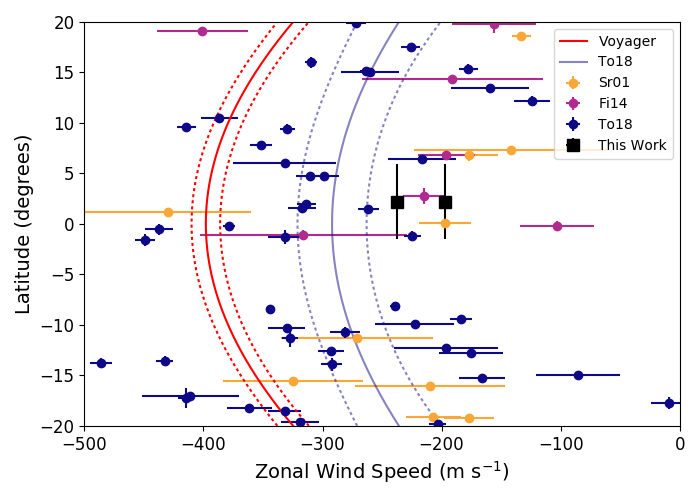}
	\caption{Drift rate of cloud features near Neptune's equator measured by various authors. The Voyager profile is the symmetric fourth-order polynomial given by \citet{sromovsky93} based on points from \citet{limaye91}. Sr01 refers to equatorial features tracked by \citet{sromovsky01b} in HST observations.  Fi14 refers to \citet{fitzpatrick14}, whose wind speeds came from Keck H-band observations. To18 refers to the Keck H-band data in \citet{tollefson18}.\label{wspeed_compare}}
\end{figure}

The size and brightness of the equatorial storm as well as the relatively high cadence of our observations permitted tracking of the storm over several months. The equatorial wind speeds of 202 m s$^{-1}$ and 237 m s$^{-1}$ we derived for the storm are compared to previous determinations of the equatorial wind speed in Figure \ref{wspeed_compare}. Our wind speeds are around a factor of two smaller than the average equatorial drift rate of $\sim$400 m s$^{-1}$ derived from Voyager spacecraft measurements at visible wavelengths (combination of green, orange, clear, and methane-U filters---see Table \ref{filtertable}) \citep[][]{limaye91, sromovsky93}, but closer to the average equatorial drift rate of $\sim$300 m s$^{-1}$ derived from Keck H-band images by \citet{tollefson18}. Both of these average wind speed fits were derived from measurements with relatively high scatter, meaning that a wind speed of 202 m s$^{-1}$ is not a clear outlier in either dataset. This can be seen in the \citet{tollefson18} points in Figure \ref{wspeed_compare}, which contain measurements with small error bars ranging from 200 to 450 m s$^{-1}$. Several other authors \citep{sromovsky01b, martin12, fitzpatrick14} have tracked equatorial features using Keck or HST observations, and all found similarly large scatter in the equatorial wind speed, with measurements ranging from 150 to 400 m s$^{-1}$. However, it is worth noting that all of the literature measurements were derived from continuous observations of small cloud features over many hours in a single night, whereas the equatorial feature was tracked occasionally over many weeks. The data are therefore sensitive to different timescales, and the average equatorial drift rate may be the most useful point of comparison to our data. \citet{tollefson18} found a significant difference in the wind speed they derived from H and Kp band measurements at the equator, which they attributed to vertical wind shear: Kp-bright features were higher in the atmosphere than Kp-faint features and were moving 90 to 140 m s$^{-1}$ more quickly on average. In contrast, the drift rate of the equatorial storm was found to be the same in H and Kp band as well as at the visible and near-IR wavelengths used by amateur observers over several months. The deeper ($\gtrsim 0.9$ bar - see Figure \ref{storm_morph}) secondary feature observed in the second epoch (after 6 October) also drifted at the same rate, providing evidence that the two features formed part of the same storm system anchored deep in the atmosphere.

Although the equatorial storm with a cloud top at 0.3-0.6 bar was always seen in association with a northern cloud complex, the pressure of the cloud top in the northern complex was significantly lower at $\lesssim$0.1 bar (see Figure \ref{storm_morph} and Section \ref{section_rt}). We take this as an upper limit on the pressure because the Kp/H ratio varied little for clouds at even lower pressures (see Figure \ref{new_ratio_plot}). The values we derived agree well with detailed spectroscopic analyses: \citet{irwin11, irwin14, irwin16} found cloud pressures of 0.1-0.2 bar at the northern midlatitudes and 0.3-0.4 bar for small equatorial ``intermediate-level'' clouds using VLT SINFONI observations. \citet{depater14} derived pressures of 0.25-0.35 bar for the two equatorial clouds in their analysis of Keck NIRC2 data, but observed mostly stratospheric ($P_m < 0.05$) bar and deep ($P_m \gtrsim 0.5$) bar clouds at the midlatitudes. \citet{gibbard03}, who observed Neptune's clouds with the NIRC2 instrument on Keck, favored stratospheric (0.02-0.06 bar) clouds at northern midlatitudes but found clouds at 0.1-0.2 bar pressures at southern midlatitudes. Interestingly, on 6 October the fainter secondary equatorial cloud was not detected at all in the Kp filter, pointing to a much deeper cloud pressure of $\gtrsim$0.9 bar.

\subsection{Dynamical Origin of the Equatorial Storm}

\subsubsection{Anticyclone Interpretation}

The different drift rate of the storms we observed with respect to the Voyager winds, the different drift rates in different epochs, and the similarity in drift rate with pressure may indicate a deep origin of the equatorial and northern clouds; the same kind of effects have been observed in convective storms in Jupiter \citep[][]{sanchezlavega08, sanchezlavega17} and Saturn \citep[][]{sanchezlavega11b}.
An upwelling region in an area of the planet predicted by circulation models to be downwelling \citep[e.g.][]{depater14} may be caused by an anticyclone-like vortex, as for the Great Dark Spot (GDS). Infrared-bright cloud features were observed along the southward edge of the southern GDS during the 1989 Voyager flyby \citep[][]{smith89}, interpreted as methane condensation produced by upwelling air from the vortex. Prominent companion clouds were also observed in association with the 1994-96 northern dark spots NGDS-15 and NGDS-32 \citep[][]{hammel97, sromovsky02}. In most images these companion clouds were only observed poleward of the dark spot; however, on 10 October 1994 a very bright, extended cloud feature was observed to the south of NGDS-32 at latitudes $<$15$^{\circ}$N.
Based on this, a companion dark spot might be expected at latitudes $\lesssim15^{\circ}$N; however, neither we (Figure \ref{hstfig}) nor \citet{wong18} observed a dark spot in the HST OPAL images taken on 6 October 2017, except at $\sim$45$^\circ$S. Therefore if the clouds we observed were supported by a deep vortex, it was either too small to be detectable by HST or was covered by the bright storm clouds even at blue wavelengths \citep[][]{hueso17, wong18}. The fact that the compact cloud remained stable in a region where the Coriolis force approaches zero, that no dark vortex is observed in HST observations in blue wavelengths at these latitudes, and that several similar clouds developed over the studied period, suggest that the bright clouds were not caused by vortices but could have been a manifestation of convective upwellling from different coherent systems in each of the bright equatorial clouds appearing in different epochs.

\subsubsection{Moist Convection Interpretation}

We explore the possibility that the bright spot was produced by moist convection. In the cloudy regions of the giant and icy planets moist air is heavier than dry air (at the same temperature) \citep[][]{guillot95, sanchezlavega04} and particular conditions are required to trigger moist convection so that latent heat release counteracts the larger density of the moist condensing air. \citep[see, e.g.,][]{hueso04}. Methane moist convection in the upper cloud of Neptune has been studied by \citet{stoker89}. We first obtain a crude estimate of the buoyancy of ascending parcels from the temperature difference between saturated updrafts (heated by methane condensation) and dry downdrafts, as given by
\begin{equation}
	\label{deltaT}
	\Delta T \approx q(L_{CH_4} / C_P)
\end{equation}
where $q$ is the methane mass mixing ratio, $L_{CH_4} = 553$ KJ kg$^{-1}$ is the latent heat of condensation of methane, and $C_P = 1310$ J kg$^{-1}$ K$^{-1}$ is the specific heat of methane at constant pressure. For a methane volume mixing ratio $f_B \approx 0.02-0.04$ and an atmospheric molecular weight $\epsilon = 6.97$, $q = \epsilon f_B \approx 0.14 - 0.28$ and Equation \ref{deltaT} gives $\Delta T \approx 6 - 12$ K. This value can be taken as an upper limit of the temperature difference between updrafts and the environment. The Convective Available Potential Energy (CAPE) is related to the peak vertical velocity $w_{max}$ reached by the updrafts \citep[][]{sanchezlavega11}:
\begin{equation}
	CAPE = \frac{w_{max}^2}{2} = \int_{Z_{f}}^{z_{n}} g \Big( \frac{\Delta T}{T} \Big) dz \approx \frac{g \Delta T}{\langle T \rangle} \Delta z
\end{equation}
where $z_f$ is the height of the free convection level, $z_n$ is the height of the equilibrium (neutral buoyancy) level, and $g = 11.1$ m s$^{-2}$ is the gravitational acceleration in Neptune's troposphere. A parcel that reaches the 0.3 bar altitude level (where cloud tops are observed) having started its motion at the 1.5 bar level within the methane cloud ($\Delta z \approx 35$ km) should have a maximum vertical velocity of 
\begin{equation}
	\label{wmax}
	w_{max} \approx \sqrt{ 2g \Big( \frac{\Delta T}{\langle T \rangle} \Delta z \Big) }
\end{equation}
which comes out to $w_{max} \sim 260-370$ m s$^{-1}$. This crude estimation agrees with the peak value in the vertical velocity profiles obtained from a one-dimensional model by \citet{stoker89} and indicates that moist convection, if initiated, can be very vigorous in Neptune.
This vertical velocity determination is probably an overestimate for several reasons. First, radiative effects near the tropopause create stable conditions that would reduce CAPE, since the temperature in the convective plume is adiabatic \citep[][]{guillot95}. Second, if convective activity is vertically confined within a limited layer, then the amount of CH$_4$ available for latent heating would be less than $q$. The temperature difference in Equation \ref{deltaT} would then be smaller. Third, including other effects in the updrafts, such as the entrainment of surrounding air on the ascending parcel, turbulent dissipation, and the weight of the condensing methane ice particles, will lower this value, limiting the altitude penetration in the atmosphere. The presence of vertical wind shears, suspected from the low velocity of the feature relative to the Voyager profile \citep[see also][]{tollefson18}, will also put serious constraints on the vertical propagation of the parcels \citep[][]{hueso04}. 
If the bright equatorial spot was convective in origin, it should have been formed by cumulus clusters with a horizontal size similar to $\Delta z$, which is $\sim$35 km. The vigorous ascent of the large number of cumulus clusters necessary to cover the whole storm area, when interacting with the sheared zonal flow, would produce the growth of a zonal disturbance, as observed in Jupiter and Saturn and whose propagation could reach the planetary scale \citep[][]{sanchezlavega11b, sanchezlavega17}. However, the images of the equatorial storm and surrounding areas did not show the presence of such a disturbance. It is possible that the disturbance occurred at a deeper cloud level than that of the top of the convective clouds, and perhaps with a lower contrast so as to be hidden at the observed wavelengths.

The energy for moist convection may also be produced by condensation of a water cloud, analogous to observed convective upwelling events in Jupiter and Saturn \citep[][]{stoker86, sanchezlavega87, gierasch00, hueso01, hueso02, hueso04}. However, this scenario is both unlikely and difficult to model accurately for the following three reasons. First, the deep oxygen and water abundances, as well as the vertical temperature profiles obtained from dry and wet adiabatic extrapolations to the depth of water condensation, are highly uncertain \citep[][]{owen06, wong08, slc13a, mousis18}. Thermochemical models \citep[][]{depater91, atreya05} predict the formation of water clouds at pressure levels $P \approx 100 - 500$ bar (depending on the deep abundance of water); that is, about 300-400 km below the observable level of 0.5-1 bar. These uncertainties mean that a simple calculation of the CAPE (Equations \ref{deltaT}-\ref{wmax}) at the deep water clouds with updrafts reaching the 0.5 bar level results in uncertain and unrealistically large vertical velocities.  Second, other cloud layers, such as NH$_4$SH, H$_2$S, and/or NH$_3$ condensates, are predicted to form in between the upper methane cloud at 0.5-1 bar and the water clouds at 100-500 bar. Updrafts that start at the water clouds and propagate across large vertical distances would interact with these cloud layers in a complicated way. Full 3-D moist convection models that include microphysics and the presence of the stacked layers of different cloud types are necessary to explore this situation, but to our knowledge these models have not yet been developed for Neptune. Third, \citet{cavalie17} have shown based on mixing length theory that H$_2$O condensation in Neptune would stabilize the atmosphere against convective motions, producing vertical gradients in water molecular weight. This is particularly important in the case of high oxygen (and therefore water) abundance, as some models predict, leading to significant temperature jumps at the water condensation layer.

\subsubsection{Wave Interpretation}

Another possible interpretation for the nature of this feature is that it formed part of an equatorial wave system. The compactness and brightness of the cloud could be related to the confinement of the moist air mass in a region showing a perturbation in the temperature, geopotential and local wind field caused by the wave. The slow motion of the feature relative to the Voyager profile could represent the zonal phase speed $c_x$ of the wave. Adopting $c_x \approx$ -202 or -237 m s$^{-1}$ and taking from Voyager profile $u \approx -400$ m s$^{-1}$ we get $c_x - u \approx 167$ or $200$ m s$^{-1}$, i.e. the spot moved eastward relative to the mean flow.
A variety of eastward and westward waves have been observed and described at the equator of the atmospheres of Venus, Earth, Mars, Jupiter and Saturn \citep[][]{allison90, sanchezlavega11, simon15a}. The simplest description of equatorial waves is in the context of the shallow-water model with linearized equations, on an equatorial $\beta$-plane for a fluid with a mean depth $h$ \citep[][]{matsuno66, andrews87, sanchezlavega11}. Three modes of eastward propagating waves result from this analysis: Rossby-gravity or Yanai (RG), inertia-gravity (IG) and Kelvin (K) modes. IG modes have been proposed to explain the temperature oscillations observed at pressures $<$1 bar during the ingress and egress at mid and high latitudes from Voyager 2 radio-occultation experiment \citep[][]{hinson93}. The horizontal (latitude-longitude) velocity structure and vertical (pressure) perturbation patterns for eastward-propagating RG and IG waves shows that they occur at both sides of the Equator with symmetric and anti-symmetric patterns, whereas for the Kelvin mode the wave perturbation patterns are centered at the equator and zonally aligned \citep[][]{matsuno66, wheeler00}. Identification of the convergence and divergence patterns \citep[][]{wheeler00} with cloud formation suggests that the observed bright Neptune spot would correspond to a Kelvin mode pattern. Due to the lack of data we cannot disregard the possibility that the feature is an RG or IG eastward mode, but here we show that these data are compatible with a Kelvin mode. Numerical simulations using the shallow water model have shown that Kelvin waves form in Jupiter's equatorial jet \citep[][]{legarreta16}. Therefore we explore the Kelvin wave as responsible for the Neptune spot. The eastward phase speed for the Kelvin wave is given by
\begin{equation}
	c_K = \sqrt{gh}
\end{equation}
and using $c_K = 167$ or 200 m s$^{-1}$ we get $h = 2.5 - 3.6$ km or about $H/6$, where $H \sim 18$ km is Neptune's atmospheric scale-height.  This Kelvin wave should be confined to a narrow atmospheric layer. In addition, the velocity and geopotential perturbation in the Kelvin wave vary with latitude as a Gaussian function centered at the equator. The e-folding decay width is given by
\begin{equation}
	y_K = \Big| \frac{2c_k}{\beta} \Big|^{1/2}
\end{equation}
and for $\beta = 2\Omega/R_N = 8.72\times10^{-12}$ m$^{-1}$ s$^{-1}$ ($\Omega = 1.8\times10^{-4}$ s$^{-1}$; $R_N = 24764$ km) we get $y_K \sim 6500$ km  which is consistent with the measured size of the bright spot. The fact that two spots were seen in some cases (see Figure \ref{data_thumbnails}) with a longitudinal separation between them about the size of the spots themselves also agrees with the horizontal structure derived for a Kelvin wave from such a model \citep[][]{matsuno66}.

\section{Summary}
\label{section_conclusions}

We have discovered a large, long-lived storm at Neptune's equator. Using near-infrared adaptive optics snapshot imaging from Keck and Lick Observatories, optical imaging from HST as part of the OPAL program, and near-infrared imaging from amateur astronomers, we tracked the evolution of the storm from June 2017 to January 2018.  Our findings can be summarized as follows:

\begin{itemize}
	\item This was the first cloud feature of its size and brightness to be observed at low latitudes on Neptune, with an H-band-derived diameter of $\sim$8500 km in the zonal direction and $\sim$7000 km in the meridional direction. Its mean latitude remained near $2^\circ$N over the course of the observations. Storm activity at the equator persisted at least from 10 June to 31 December 2017 ($\gtrsim$7 months); a discrete feature was observed from 26 June to 25 July but had broken up into a trail of small clouds by 04 September.  On 4 October a new storm had appeared, and another breakup into small cloud features was observed on 29 November.
	
	\item Feature tracking found best-fit drift rates of 201.7 $\pm$ 2.2 m s$^{-1}$ between 7 and 14 July and 237.4 $\pm$ 0.10 m s$^{-1}$ between 28 September and 4 November for the storm feature. The feature was found to vary in speed from 10 June to 25 July. The same wind speed was measured at different wavelengths, pointing to a coherent storm system anchored at $\gtrsim$1 bar pressure.
	
	\item Radiative transfer modeling suggested a cloud top pressure of 0.3-0.6 bar for the equatorial storm and $\lesssim$0.1 bar for the northern cloud complex for all four Keck observations in which the storm was observed. A decrease in reflectivity of both the northern and equatorial clouds between 26 June and 25 July was interpreted as a decrease in opacity of the clouds between the two dates.
	
	\item A secondary equatorial storm feature was observed $\sim$50$^{\circ}$ longitude away from the main storm and maintained the same drift rate as the main storm from 6 October to at least 4 November. However, the secondary feature was undetected in the Kp filter, meaning its cloud top was at $>0.9$ bar pressure, much deeper than the main storm.
	
	\item No ``dark-spot'' vortex was observed near the equator in Hubble images. The upwelling that presumably underlay the storm may therefore have been driven by a Kelvin wave or by moist convection. However, the dynamics of this rare event have yet to be studied in detail.
	
\end{itemize}

\software{Astropy, cython, emcee, image\_registration, matplotlib, nirc2\_reduce, numpy, pydisort, scikit-image, scipy, WinJupos}

\acknowledgements

{\large\textit{Acknowledgements:}} Some of the data presented herein were obtained at the W. M. Keck Observatory, which is operated as a scientific partnership among the California Institute of Technology, the University of California and the National Aeronautics and Space Administration. The Observatory was made possible by the generous financial support of the W. M. Keck Foundation.

The authors wish to recognize and acknowledge the very significant cultural role and reverence that the summit of Maunakea has always had within the indigenous Hawaiian community.  We are most fortunate to have the opportunity to conduct observations from this mountain.

Partial support for this work was also provided by the Keck Visiting Scholar Program at W.M. Keck Observatory.

Portions of this research are based on observations made with the NASA/ESA Hubble Space Telescope, (OPAL program GO14756) obtained from the data archive at the Space Telescope Science Institute. STScI is operated by the Association of Universities for Research in Astronomy, Inc. under NASA contract NAS 5-26555

Research at Lick Observatory is partially supported by a generous gift from Google.

This work has been supported in part by the National Science Foundation, NSF Grant AST-1615004 to UC Berkeley.

R. H. and A.S-L. were supported by the the Spanish MINECO project AYA2015-65041-P with FEDER, UE support and Grupos Gobierno Vasco IT- 765- 13.

Portions of this work were performed under the auspices of the U.S. Department of Energy by Lawrence Livermore National Laboratory under Contract DE-AC52-07NA27344.

We thank the referees, Amy Simon and one anonymous person, for their insightful comments, which substantially improved the manuscript.

We thank Conor McPartland as well as all of the Keck Observing Assistants for executing our volunteer observing program during their observing time at Keck Observatory.

We thank Geoff Chen, Ian Crossfield, Donald Gavel, and Robert de Rosa for executing our volunteer observing program during their observing time at Lick Observatory.

\appendix

\setcounter{table}{0}
\renewcommand{\thetable}{A\arabic{table}}
\setcounter{figure}{0}
\renewcommand{\thefigure}{A\arabic{figure}}

\section{Proteus Albedo Determination}
\label{appendix_proteus}

We determined the total flux of Proteus using a flux bootstrapping method similar to \citet{gibbard05}. Since Proteus was at relatively low signal-to-noise in a background containing considerable scattered light from Neptune, only the flux from the inner core of its point spread function (PSF) could be reliably measured. In order to account for the missing flux in the PSF sidelobes, we measured the flux $F_{0.2}$ from a $0.2\arcsec$ radius aperture around the standard star, and compared that with the flux from a large aperture containing all the flux $F_{tot}$ from the star. On 25 July a $1.0\arcsec$ radius aperture was used to compute $F_{tot}$, and a $0.5\arcsec$ aperture was used on 26 June. This difference was due to the better atmospheric seeing on 26 June: the bright core of the PSF from HD1160 forced the use of the 128 pixel subarray to avoid saturating the NIRC2 detector, making the field of view too small to use a $1.0\arcsec$ radius aperture. However, since the PSF was very sharp on 26 June, a $0.5\arcsec$ radius aperture was sufficient to capture all of the flux. The ratio $F_{0.2}/F_{tot}$ is given in Table \ref{moonphot} for each date and band. This correction was then applied to the measured flux from Proteus. The total flux from the moon was also divided by the moon's projected surface area to retrieve an I/F value. We assumed Proteus had a spherical shape with a 210$\pm$7 km radius \citep[][]{karkoschka03}, which corresponded to $\approx$0.01$\arcsec$ at our pixel scales of 213 km px$^{-1}$ on 26 June and 210 km px$^{-1}$ on 25 July; this was smaller than Keck's diffraction limit of $\approx$0.04$\arcsec$ at 1.6 $\mu$m, so the moon was unresolved. The final I/F value was simply the geometric albedo and is given in Table \ref{moonphot}, ignoring phase angle effects. The phase angle was 1.8$^\circ$ on 26 June and 1.2$^\circ$ on 25 July, small enough that Proteus's surface was in near full sun but large enough that coherent backscattering was not yet important \citep[][]{karkoschka01}. The additional error introduced by our flux bootstrapping technique was estimated by varying the inner aperture size from 0.15$\arcsec$ to 0.3$\arcsec$ and recalculating the final I/F value for each; the error was found to be 5-10\% except in the Kp filter on 25 July, for which the error was 26\% due to variable atmospheric seeing. The bootstrapping error was added in quadrature to our 20\% standard star photometric error, and the total is shown in Table \ref{moonphot}.

\section{Supplementary Data}

\begin{figure}
	\includegraphics[width = 1.0\textwidth]{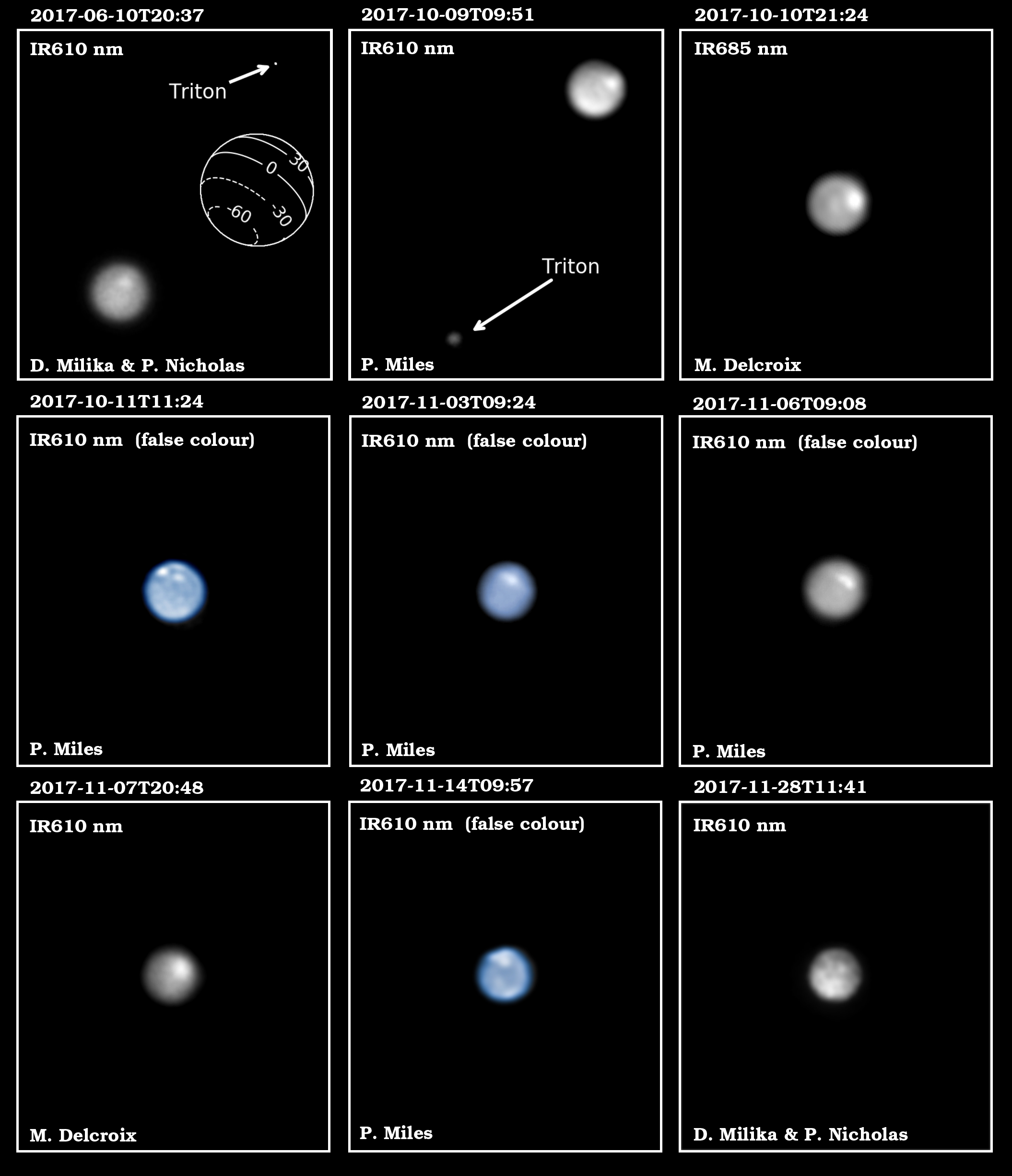}
	\caption{Selection of Neptune observations with small telescopes by different observers. Each panel covers a field of view of 29 $\times$ 32 arcsec; the orientation is sky North up and sky East to the left in all images. The sequence shows the initial equatorial feature (upper row) since its first observation (upper left panel), and the multiple features observed over November 2017 (middle and bottom rows). All of the images had Triton visible in their original field of view, and the top center panel is offset so Triton can be seen. Individual observers and filters are identified in each panel.\label{amateur_appendix}}
\end{figure}

Table \ref{amateurdata} contains information about the amateur and PlanetCam observations used in this paper, and Figure \ref{amateur_appendix} shows sample thumbnail images from these observations. Table \ref{filtertable} shows central wavelengths and full bandpass widths for all filters referenced in this paper.

\section{Wind Speed Retrievals}
\label{appendix_tracking}

The longitude tracking data were fit to a linear wind speed via Markov Chain Monte-Carlo (MCMC) maximum likelihood estimation, implemented by the \texttt{emcee} package in \texttt{Python} \citep[][]{dfm13}\footnote{http://dfm.io/emcee/current/}. Our application of this technique is explained briefly here.

Assuming a cloud on Neptune drifts at a linear wind speed $w$, and given a time $t_0$ at which the cloud's longitude is $L_0$, the longitude $L_m$ of the cloud at any other time $t_n$ is given by
\begin{equation}
	\label{modelwind}
	L_m(w, t_n, L_0, t_0) = L_0 + (t_n - t_0)w
\end{equation}
The likelihood function $\ln p$ is then
\begin{equation}
	\label{lnlike}
	\ln p(L | t, \sigma, w, f) = -\frac{1}{2} \sum_n \Big[ (L_n - L_m(w, t_n, L_0, t_0))^2 s_n^{-2} + \ln (s_n^{-2}) \Big]
\end{equation}
where $s_n^2 = \sigma_n^2 + f^2$ is the longitude variance. Writing the variance this way allows for the possibility that the longitude error $\sigma_n$ (see Section \ref{section_tracking} for an explanation of how this was determined) was underestimated by some constant amount $f$. The MCMC algorithm maximizes $\ln p$: in each step of the retrieval, the algorithm chooses values of $w$ and $f$, uses $w$ to predict longitudes $L_m$ at each time $t_n$ according to Equation \ref{modelwind}, evaluates how well the longitude data are fit by that model using Equation \ref{lnlike}, and then chooses a new $w$ and $f$ pair based on the goodness of fit.

\begin{figure}
	\includegraphics[width = 0.5\textwidth]{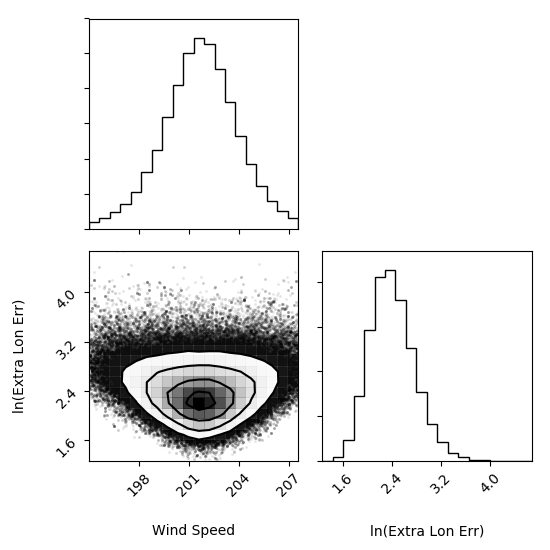}
	\includegraphics[width = 0.5\textwidth]{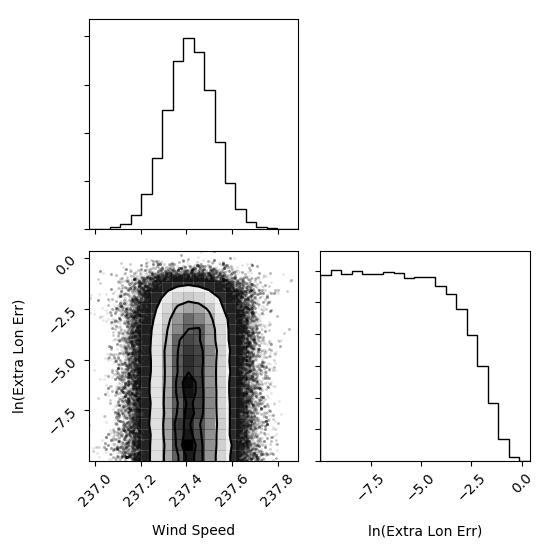}
	\caption{``Corner plots'' showing the one- and two-dimensional projections of the posterior probability distributions of the MCMC-retrieved parameters for the wind speed retrievals on \textbf{Left:} the first epoch of observations, using only the data from 2 to 14 July (where the observations are sampled most densely) and \textbf{Right:} the second epoch of observations from 30 September to 26 October.\label{corner}}
\end{figure}

Retrievals were carried out separately for the two epochs of observation. Corner plots for both retrievals are shown in Figure \ref{corner}. The retrieval favored nearly Gaussian errors on the wind speed in both epochs, with best-fit values of $201.7 \pm 2.2$ and $237.4 \pm 0.2$, respectively. An additional longitude error of $\sim$11$^{\circ}$ was favored by the retrieval in the first epoch (which should be added in quadrature with the original errors), but no additional error term was prescribed in the second epoch. This difference has two possible explanations. First, since half of the measurements in the first epoch were from Lick data whereas most of the measurements in the second epoch were from small telescopes, it is possible that the errors on the Lick data were underestimated while the small telescope errors were correct. The longitude errors from Lick could be underestimated due to morphological changes in the storm cloud: since Lick resolved the large cloud, the technique used to obtain the storm location actually measured the location of the brightest region of the cloud, which may have changed relative to the rest of the storm over time. The second explanation is simply that a linear wind speed fit did not adequately describe the data over the first epoch; that is, the storm's drift rate was changing over timescales of a few days from 2 to 14 July. This is reasonable because the drift rate certainly changed from 26 June to 25 July as can be seen by the large residuals in the wind speed fit shown in the first panel of Figure \ref{lon_tracking}.

\clearpage
\bibliography{paper}{}

\newpage
\begin{longtable}{|c|c|c|}


\caption{Description of amateur observations. The filter wavelengths are given in Table \ref{filtertable}.\label{amateurdata}} \\

UT Date \& Start Time & Observer & Filters \\
\hline
2017-06-10 19:45 & Darryl Milika \& Pat Nicholas &  IR610\\ 
2017-07-11 01:53 & PlanetCam$:$ Hueso, Ordonez & RG1000\\ 
2017-07-11 04:03 & PlanetCam$:$ Hueso, Ordonez & M2\\ 
2017-07-14 19:22 & Darryl Milika \& Pat Nicholas & IR610\\ 
2017-09-30 13:03 & Darryl Milika \& Pat Nicholas & IR610\\ 
2017-10-04 03:38 & Steve Fugardi   & IR610\\ 
2017-10-09 09:51 & Phil Miles  & IR610\\ 
2017-10-09 10:23 & Phil Miles  & IR610\\ 
2017-10-09 11:13 & Phil Miles  & IR610\\ 
2017-10-10 21:23 & Marc Delcroix &  IR685\\ 
2017-10-11 11:03 & Phil Miles  & IR610\\ 
2017-10-11 11:24 & Phil Miles  & IR610\\ 
2017-10-12 22:40 & Martin Lewis &  IR610\\ 
2017-10-13 02:49 & Steve Fugardi &  IR610\\ 
2017-10-13 20:30 & Lucien Polongini  & IR610\\ 
2017-10-14 12:24 & Darryl Milika \& Pat Nicholas & IR610\\ 
2017-10-15 21:30 & Martin Lewis & IR610\\ 
2017-10-17 11:07 & Darryl Milika \& Pat Nicholas & IR610\\ 
2017-10-18 21:33 & Emmanuel Kardasis & IR610\\ 
2017-10-19 17:53 & Dimitris Kolovos   & IR610\\ 
2017-10-20 11:10 & Phil Miles  & IR610\\ 
2017-10-20 11:45 & Darryl Milika \& Pat Nicholas & IR610\\ 
2017-10-21 00:59 & Antonio Checco &  IR610\\ 
2017-10-21 03:13 & Steve Fugardi  & IR610\\ 
2017-10-21 20:30 & Dimitri Kolovos   & IR610\\ 
2017-10-21 20:19 & Emmanuel Kardasis &  IR610\\ 
2017-10-25 13:04 & Phil Miles  & IR610\\ 
2017-10-26 04:36 & Blake Estes  & IR685 \\ 
2017-10-26 05:37 & Randy Christensen   & IR610\\ 
2017-10-30 17:14 & Clyde Foster &   IR610\\ 
2017-10-30 17:48 & Clyde Foster &   IR610\\ 
2017-10-31 09:35 & Phil Miles  & IR610\\ 
2017-10-31 11:29 & Darryl Milika \& Pat Nicholas  & IR610\\ 
2017-10-31 11:59 & Darryl Milika \& Pat Nicholas  & IR610\\ 
2017-10-31 12:23 & Darryl Milika \& Pat Nicholas  & IR610 \\
2017-11-01 21:19 & Martin Lewis   & IR610\\
2017-11-02 18:54 & Nick Haigh   & IR \\  
2017-11-03 09:23 & Phil Miles   & IR610\\ 
2017-11-03 10:26 & Phil Miles   & IR610\\ 
2017-11-04 19:57 & Clyde Foster  & IR610\\ 
2017-11-04 20:10 & Clyde Foster  & IR610\\ 
2017-11-04 20:58 & Clyde Foster  & IR610\\ 
2017-11-06 09:08 & Phil Miles &  IR610\\ 
2017-11-06 10:11 & Phil Miles &  IR610\\ 
2017-11-07 18:36 & John Sussenbach   & IR685\\ 
2017-11-07 18:44 & John Sussenbach   & IR685\\ 
2017-11-07 19:20 & Manos Kardasis   & IR610\\ 
2017-11-07 20:00 & Michel Miniou &   IR \\ 
2017-11-07 20:48 & Marc Delcroix   & IR610\\ 
2017-11-08 11:07 & Phil Miles &   IR610\\ 
2017-11-08 11:36 & Darryl Milika \& Pat Nicholas &   IR610\\ 
2017-11-08 11:42 & Phil Miles   & IR610\\ 
2017-11-08 12:25 & Phil Miles   & IR610\\ 
2017-11-10 18:51 & Clyde Foster    & IR685\\ 
2017-11-11 09:48 & Phil Miles   & IR610\\ 
2017-11-11 11:14 & Phil Miles   & IR610\\ 
2017-11-11 12:11 & Phil Miles   & IR610\\ 
2017-11-13 18:04 & John Sussenbach    & IR685\\ 
2017-11-14 09:55 & Anthony Wesley    & IR610\\ 
2017-11-14 09:56 & Phil Miles &    IR610\\ 
2017-11-14 10:08 & Phil Miles &    IR610\\ 
2017-11-14 10:27 & Phil Miles &    IR610\\ 
2017-11-14 10:54 & Phil Miles &    IR610\\ 
2017-11-15 23:30 & Almir Germano   & IR610 \\
\end{longtable}

\begin{table}
	\footnotesize
	\begin{tabular}{|c|c|c|c|}
Instrument & Filter & $\lambda_c$ ($\mu$m) & $\Delta \lambda$ ($\mu$m) \\  
\hline 
Keck NIRC2 & H & 1.63 & 0.30 \\ 
Keck NIRC2 & Kp & 2.20 & 0.35 \\ 
Keck NIRC2 & CH4S & 1.59 & 0.13 \\ 
Keck NIRC2 & PaBeta & 1.290 & 0.019 \\ 
Lick ShARCS & H & 1.66 & 0.30 \\ 
Lick ShARCS & Ks & 2.15 & 0.32 \\
HST WFC3 & F467M & 0.4675 & 0.0230 \\ 
HST WFC3 & F547M & 0.5475 & 0.0710 \\ 
HST WFC3 & FQ619N & 0.6194 & 0.0062 \\ 
HST WFC3 & F657N & 0.6573 & 0.0094 \\ 
HST WFC3 & F763M & 0.7630 & 0.0780 \\ 
HST WFC3 & F845M & 0.8454 & 0.0870 \\ 
Calar Alto PlanetCam & RG1000 & 1.0 & 0.6 \\
Calar Alto PlanetCam & M2 & 0.727 & 0.005 \\
Amateur & IR610 & 0.610 & LP \\
Amateur & IR685 & 0.685 & LP \\
Voyager ISS & Clear & 0.460 & 0.360 \\
Voyager ISS & Green & 0.585 & 0.110 \\
Voyager ISS & Orange & 0.615 & 0.050 \\
Voyager ISS & Methane-U & 0.540 & 0.012
	\end{tabular}
	\caption{Central wavelengths and full bandpass widths for Keck, Lick, HST, PlanetCam, amateur, and Voyager filters referenced in this paper. ``LP'' denotes a long-pass filter.\label{filtertable}}
\end{table}

\end{document}